\documentclass[prd,preprint,superscriptaddress,amsmath,amssymb]{revtex4}

\usepackage{graphicx,color,slashed}

\begin{document}

{\begin{flushright}{KIAS-18101}
\end{flushright}}

\title{\bf  Left-handed color-sextet diquark in the Kaon system}

\author{Chuan-Hung Chen}
\email{physchen@mail.ncku.edu.tw}
\affiliation{Department of Physics, National Cheng-Kung University, Tainan 70101, Taiwan}

\author{Takaaki Nomura}
\email{nomura@kias.re.kr}
\affiliation{School of Physics, KIAS, Seoul 02455, Korea}

\date{\today}

\begin{abstract}
We investigate whether a color-sextet scalar diquark (${\bf H}_6$) coupling to the left-handed quarks contributes to the $\Delta S=2$ process. It is found that the box diagrams mediated by $W$ and ${\bf H}_6$ bosons have no contributions to $\Delta S=2$ when the limit of $m_t=0$ is used, and the flavor mixing matrices for diagonalizing quark mass matrices are introduced at the same time. When the heavy top-quark mass effects are taken into account, it is found that in addition to the  $W-{\bf H}_6$ box diagrams significantly contributing to $\Delta S=2$,  their effects can be as large as those from the ${\bf H}_6-{\bf H}_6$ box diagrams. 
Using  the  parameters that are constrained by the $K^0-\bar K^0$ mixing parameter $\Delta M_K$ and the Kaon indirect CP violation $\epsilon_K$, we find that the left-handed color-sextet diquark can lead to the Kaon direct CP violation being $Re(\epsilon'/\epsilon) \sim 0.3 \times 10^{-3}$. In the chosen scheme, although the diquark contribution to $K_L\to \pi^0 \nu \bar\nu$ is small, the branching ratio of $K^+ \to \pi^+ \nu \bar\nu$ can reach the current experimental upper bound. 
\end{abstract}

\maketitle

\section{Introduction}

Despite the success of the standard model (SM) in explaining most experimental data, the SM is an effective theory only at the electroweak (EW) scale because some  long standing phenomena are still puzzling, such as  baryogenesis, neutrino mass, and the muon anomalous magnetic dipole moment. More recently,  from  the RBC-UKQCD collaboration~\cite{Blum:2015ywa,Bai:2015nea} using the lattice QCD and the group using a large $N_c$ dual QCD~\cite{Buras:2015xba,Buras:2015yba,Buras:2016fys},  it was found that  the predicted Kaon direct CP violation in the SM  is less than  the experimental data by more than a $2\sigma$. If this inconsistency is  finally confirmed, it indicates the necessity of a new physics effect. 

A new mechanism with a colored scalar (e.g., diquark), which was used to resolve the strong CP problem and can contribute to the Kaon indirect ($\epsilon_K$) and direct $(\epsilon'/\epsilon)$ CP violation, was originally proposed in~\cite{Barr:1986ky,Barr:1989fi}. Although a diquark can originate from grand unified theories (GUTs)~\cite{Barr:1986ky,Assad:2017iib}, it can be an unobserved particle with a mass of order of  TeV in the SM gauge symmetry $SU(3)_C\times SU(2)_L \times U(1)_Y$. 

If we take  the diquark as the particle in the SM gauge symmetry, the possible representations can be: $(3,1,-1/3)$, $(6,1,1/3)$, $(3,3,-1/3)$, and $(3,1,2/3)$~\cite{Barr:1989fi}, where $(3,1,-1/3)$ and $(6,1,1/3)$ can couple to the $SU(2)_L$ quark doublets and singlets. Moreover, it can be verified that before EW symmetry breaking (EWSB), the Yukawa couplings of ${\bf H}_3=(3,1,-1/3)$ and ${\bf H}_6=(6,1,1/3)$ to the  left-handed quark doublets are flavor symmetric and antisymmetric, respectively. Although the ${\bf H}_3$ Yukawa coupling matrix in flavor space is symmetric, there are six independent Yukawa matrix elements; thus, the symmetric property, similar to a generic case, may not exhibit unique behavior in the flavor physics. The detailed analysis of ${\bf H}_3$ can be found in~\cite{Chen:2018dfc}. However, the antisymmetric ${\bf H}_6$ Yukawa matrix only has three independent elements; thus, if we assume that the new Yukawa couplings are real parameters, there are only four  new parameters involved, including the diquark  mass.  

In addition to involving fewer parameters, ${\bf H}_6$ has some interesting characteristics.  It was argued in~\cite{Barr:1989fi} that the box diagrams mediated by one $W$ gauge boson and one ${\bf H}_6$ for $\Delta S=2$  vanish. We revisit the issue and find that the conclusion is only correct in the limit of $m_t=0$; here, we still  take the light quarks to be massless, i.e. $m_{u,c}=0$. Moreover, to achieve the vanished result, the unitary  flavor mixing matrices introduced  to diagonalize quark mass matrices have to be  simultaneously included. The situation is very different from the case of ${\bf H}_3$, where due to the antisymmetric property in color space, the same box diagrams using ${\bf H}_3$ instead of ${\bf H}_6$ vanish, even when the $m_t$ effect is taken into account~\cite{Barr:1989fi,Chen:2018dfc}. 

In order to show the importance of the heavy top-quark mass effect, we study the $\Delta S=2$ processes with and without the limit of $m_t=0$ in detail.   For the massive top-quark, in addition to the $W$-mediated box diagrams, we have to consider the charged-Goldstone-boson $(G)$ contributions when the 't Hooft-Feynman gauge is used. It is found that the contribution  to $\epsilon_K$ from  one $W(G)$ and one ${\bf H}_6$ box diagrams can be as large as that from the pure diquark box diagrams, which  are insensitive to $m_t$.  

 When the  parameters in the ${\bf H_6}$ model are constrained by the  $K^0-\bar K^0$ mixing parameter $\Delta M_K$ and the indirect CP violation $\epsilon_K$, we consider the implications on $\epsilon'/\epsilon$ and $K\to \pi \nu \bar\nu$. It is found that if we take the antisymmetric ${\bf H}_6$ Yukawa matrix to be real, $\epsilon'/\epsilon$ from the diquark contribution can reach  a  value of around $0.3 \times 10^{-3}$, which is comparable to the SM result. Although there is no new CP violating effect contributing to  $K_L \to \pi^0 \nu \bar\nu$, we find that the branching ratio (BR) for the $K^+ \to \pi^+ \nu \bar\nu$ decay induced by the ${\bf H}_6$-mediated $Z$-penguin  can reach the upper limit of the current experimental data. The NA62 experiment at CERN plans to measure  $BR(K^+ \to \pi^+ \nu \bar\nu)$ with  a $10\%$ precision~\cite{Rinella:2014wfa,Moulson:2016dul}. If an unexpected large BR for $K^+ \to \pi^+ \nu \bar\nu$ is found in NA62, the diquark may be a potential candidate explaining the anomaly. On the other hand, if  the measured $BR(K^+\to \pi^+ \nu \bar\nu)$ is close to the SM result, then we can use it to bound the parameter space of the diquark coupling. 

The paper is organized as follows: We introduce the Yukawa couplings and EW gauge couplings of the color-sextet diquark in Sec. II. We calculate the diquark box diagrams for $\Delta S=2$ with and without the limit of $m_t=0$ in Sec. III. In addition, assuming that the ${\bf H}_6$ Yukawa matrix is a real matrix,   we  study the  $\epsilon_K$ and $\Delta M_K$ constraints on the free parameters. In Sec. IV,  using the constrained parameters, we study the diquark contributions to the $\epsilon'/\epsilon$ and $K\to \pi \nu \bar\nu$.   A summary is given in Sec. VI.

\section{Color-sextet diquark Yukawa and gauge couplings}

In order to study the color-sextet diquark effects, in this section, we introduce  the diquark Yukawa  couplings to the SM quarks and  its  gauge couplings to the EW gauge bosons. With the real diquark Yukawa  matrix,  the loop contributions to $\epsilon'/\epsilon$ and $K\to \pi \nu \bar\nu$ are dominated by the $Z$-penguin~\cite{Chen:2018dfc}, so, we skip discussions of the diquark coupling to the gluons.

\subsection{Yukawa couplings}

 The SM gauge invariant   Yukawa couplings of ${\bf H}_6(6,1,1/3)$ to the left-handed quarks can be written as:
 \begin{equation}
 -{\cal L}_{Y} =  Q^T C{\bf y} \pmb{\varepsilon} {\bf H}^\dagger_{6} P_L Q + H.c, \label{eq:L_Y}
 \end{equation}
 where  the flavor indices are suppressed; ${\bf y}$ is the Yukawa matrix, and $\pmb{\varepsilon}$ is a $2\times 2$ antisymmetric matrix with  $\pmb{\varepsilon}_{12}=-\pmb{\varepsilon}_{21}=1$. The ${\bf H}_6$ representation  in $SU(3)_C$ can be expressed as ${\bf H}_{6} = S^a H^a_{6}$ ($a=1$-$6$), where the  symmetric matrix forms of $S^a=(S^a)^T$ can be found in~\cite{Han:2009ya}. For the complex conjugate state, we use $(\bar S_a)_{\alpha\beta}=(S^a)^{\beta \alpha}$, i.e.  ${\bf H}^\dagger_6= \bar S_a H^{*}_{6a}$; thus, we obtain $Tr (S^a \bar S_b)=\delta^a_b$ and $(S^a)^{\beta \alpha} (\bar S_a)_{\rho\sigma}=1/2(\delta_\sigma^{\beta}  \delta_\rho^{\alpha}+\delta_\rho^{\beta}\delta_\sigma^{\alpha})$. Using the fermion anticommutation relations, the antisymmetric $C$ and $\pmb{\varepsilon}$, and the  symmetric $P_L$ and ${\bf H}_6$, we can verify that  ${\bf y}$ is an  antisymmetric matrix, e.g., ${\bf y}^T = - {\bf y}$. Since our purpose is to study the antisymmetric effects of ${\bf y}$, we will concentrate on the ${\bf H}_6$ couplings to the left-handed quarks and we briefly discuss the effect of couplings to the right-handed quarks in Sec.~\ref{sec:right-handed}. 
  
  Basically, we can use ${\bf y}$ to investigate the relevant phenomena~\cite{Barr:1989fi}; however, the flavor mixings induced from the electroweak symmetry breaking (EWSB) may change the antisymmetric property of the Yukawa matrix that originally arises from  ${\bf y}$. To see the influence of the quark-flavor mixings, we introduce the unitary matrices $V^{u,d}_L$ to diagonalize the quark mass matrices, and  Eq.~(\ref{eq:L_Y}) in terms of quark mass eigenstates can then be expressed as:
  \begin{equation}
   -{\cal L}_{Y} 
    = u^T_L C \left[ \left( V^{u*}_L {\bf y}  V^{d\dagger}_L\right) -  \left( V^{d*}_L {\bf y}  V^{u\dagger}_L\right)^T \right] {\bf H}^\dagger_{6}  d_L+ H.c.
     \end{equation}
   Using $\left( V^{d*}_L {\bf y}  V^{u\dagger}_L\right)^T=- V^{u *}_L {\bf y} V^{d\dagger}_L$, we obtain:
   \begin{equation}
    -{\cal L}_{Y} = u^T_L C {\bf g} {\bf H}^\dagger_{6}  d_L+ H.c.\,, ~ {\bf g}= 2 V^{u *}_L {\bf y} V^{d\dagger}_L\,. \label{eq:LY}
   \end{equation}
   It can be seen that when the flavor mixings are introduced, the new Yukawa matrix ${\bf g}$ generally is not an antisymmetric matrix. If we link $V^{u,d}_L$ to the Cabibbo-Kobayashi-Maskawa (CKM) matrix, defined by $V \equiv V^u_L V^{d\dagger}_L$, the new Yukawa matrix ${\bf g}$ can be expressed as:
    \begin{equation}
    {\bf g}=2 {\bf y} V\,, \label{eq:g}
    \end{equation}
  where  $V^{u}_L =1$ is taken. According to the Wolfenstein's parametrization~\cite{Wolfenstein:1983yz}, the off-diagonal elements of the CKM matrix can be parametrized using a small parameter of $\lambda\approx 0.225$. Based on the characteristic of the CKM matrix, we can understand from the phenomenological viewpoint if ${\bf g}$ can be approximately regarded as  ${\bf y}$. 
  
  In order to understand the  difference between ${\bf y}$ and ${\bf g}$, we first analyze the diagonal elements of ${\bf g}$ as:
   \begin{align}
   g_{11}& =  2y_{12} V_{21} + 2 y_{13} V_{31}\sim - 2 y_{12} \lambda \,, \nonumber \\
   g_{22} & = 2 y_{21} V_{12} + 2 y_{23} V_{32} \sim  -  2 y_{12} \lambda \,, \nonumber \\
   g_{33} & = 2 y_{31} V_{13} + 2 y_{32} V_{23} \sim 0\,,
   \end{align}
 where we have dropped  the $\lambda^2$ and $\lambda^3$ terms that are  from the CKM matrix elements  as the approximation. If $|y_{12}| \sim  |y_{31} |\lambda^2$ or $ |y_{12}| \sim |y_{32}| \lambda$ is satisfied, we can take $g_{11}\sim g_{22}\sim 0$. Under these circumstances, $g_{ii}\approx 0$ is similar to $y_{ii}=0$.  
 Using the same approximation, we can analyze the off-diagonal elements of ${\bf g}$ as:
  \begin{align}
  g_{12} & = 2 y_{12}V_{22} + 2 y_{13} V_{32} \sim 2  y_{12} - 2 \lambda^2 y_{13} \,, ~ g_{21}  = 2 y_{21}V_{11} +2  y_{23} V_{31} \sim 2 y_{21} \,, \nonumber \\
  g_{13} & = 2 y_{12} V_{23} + 2 y_{13} V_{33} \sim 2 y_{13} \,, ~ g_{31} = 2 y_{31} V_{11} + 2  y_{32} V_{21} \sim 2 y_{31} - 2 y_{32} \lambda \nonumber \\ 
  g_{23} & = 2 y_{21} V_{13} + 2 y_{23} V_{33} \sim 2 y_{23} \,, ~ g_{32} = 2 y_{31}V_{12} + 2 y_{32} V_{22} \sim  2y_{31} \lambda + 2 y_{32}\,. \label{eq:off_g}
  \end{align}
 From Eq.~(\ref{eq:off_g}), it can be clearly seen that we cannot simply obtain $g_{ij} \approx - g_{ji}$, which is possessed by $y_{ij}$.  

\subsection{EW gauge couplings to ${\bf H}_{6}$}

 In order to obtain the photon and $Z$-boson gauge couplings to the ${\bf H}_6$ diquark, we write the $U(1)_Y$ covariant derivative of ${\bf H}_6$ as:
 \begin{equation}
 D_\mu {\bf H}_6 = (\partial_\mu + i g' Y_{H_6} B_\mu ){\bf H}_6\,,
 \end{equation}
 where $g'$ is the $U(1)_Y$ gauge coupling constant; $Y_{H_6}$ is the ${\bf H}_6$ hypercharge, and $B_\mu$ is the $U(1)_Y$ gauge field. Since ${\bf H}_6$ is an $SU(2)$ singlet, its hypercharge is equal to its electric charge, i.e., $Y_{H_6}=e_{H_6}$. 
Using $B_\mu = \cos\theta_W A_\mu - \sin\theta_W Z_\mu$, the EW  gauge couplings to the diquark can be obtained from the $U(1)_Y$ gauge invariant kinetic term of ${\bf H}_6$ and can be written as:
  \begin{align}
  {\cal L}_{VH_6 H_6}  & = i e_{H_6} e (\partial_\mu H^*_{6a} H^a_{6} - H^*_{6a} \partial_\mu H^a_6 )A^\mu \nonumber \\
  & -i  \frac{g e_{H_6} \sin^2\theta_W}{\cos\theta_W} (\partial_\mu H^*_{6a} H^a_{6} - H^*_{6a} \partial_\mu H^a_6 )Z^\mu\,, \label{eq:H6-AZ}
  \end{align} 
  where $\theta_W$ is the Weinberg's angle; $e=g'\cos\theta_W = g\sin\theta_W$ and $g'/g=\tan\theta_W$ are applied; $g$ is the $SU(2)_L$ gauge coupling constant, and $e_{H_6}=Y_{H_6}=1/3$ is the $H^a_{6}$ electric charge. The associated Feynman rules can be obtained as: 
 \begin{align}
 A_\mu  H^*_{6a} H^b_6 & : -i e_{H_6} e ( p_b + p_a)_\mu  \delta^b_a \,, \label{eq:photonH3H3}   \\
 Z_\mu H^*_{6a} H^b_6  & : i \frac{g e_{H_6} \sin^2\theta_W}{\cos\theta_W} ( p_b + p_a)_\mu  \delta^b_a \,. \label{eq:ZH3H3} 
 \end{align}

\section{${\bf H}_6$-mediated box diagrams for $\Delta S=2$}

\begin{figure}[phtb]
\includegraphics[scale=0.75]{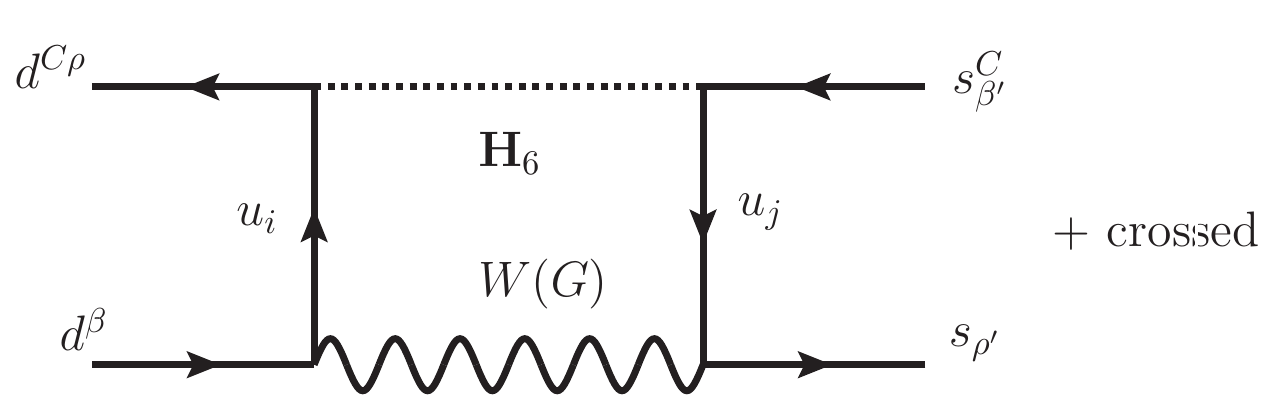}
 \caption{ Box diagram mediated by one $W(G)$ and one ${\bf H}_6$ for $\Delta S=2$, where $G$ denotes the charged-Goldstone boson. }
\label{fig:Box_WH6}
\end{figure}

Based on the diquark Yukawa couplings in Eq.~(\ref{eq:LY}), we discuss the diquark effects on the $\Delta S=2$ process, where the Feynman diagram arises from the intermediates of $W(G)$ and ${\bf H}_6$ are sketched in Fig.~\ref{fig:Box_WH6}. The other types of box diagrams can be obtained using ${\bf H}_6$ instead of $W(G)$. In order to understand the effects of the massive top-quark, we distinguish the $m_t=0$ case from the $m_t\neq 0$ case.

\subsection{In the limit of $m_t = 0$}

Using the ${\bf H}_6$ Yukawa couplings in Eq.~(\ref{eq:LY}), the effective Hamiltonian for $\Delta S=2$ from one $W$ and one ${\bf H}_6$ box diagram shown in Fig.~\ref{fig:Box_WH6} can be written as:
 \begin{align}
 {\cal H}^{WH_6}_{\rm Box} & = - i \frac{g^2}{4} \sum_{i,j}\left(V^*_{j s} V_{i d} \, g^*_{j2} g_{i1}\right) (\bar S_{a} )_{\rho \beta} (S^a)^{\rho' \beta'} \int \frac{ d^4 q}{(2\pi)^4} \frac{q^{\mu'} q^{\nu'}}{(q^2 -m^2_{H_6}) (q^2 -m^2_W) q^4}  \nonumber \\
 & \times  \left( \overline{d^C}^\rho \gamma_{\mu'} \gamma_{\mu} P_L d^\beta \right) \left( \bar s_{\rho'} \gamma^\mu \gamma_{\nu'} P_R s^C_{\beta'} \right)\,, 
 \end{align}
 where $i, j$ denote the flavor indices of up-type quarks. Due to the limit of $m_{u,c,t}=0$, the charged-Goldstone-boson has no contributions to the $\Delta S=2$ process.  
 Using $\gamma_\mu \gamma_\nu = g_{\mu \nu} - i\sigma_{\mu \nu}$ and the loop integral
 \begin{equation}
 \int \frac{ d^4 q}{(2\pi)^4} \frac{q^{\mu'} q^{\nu'}}{(q^2 -m^2_{H_6}) (q^2 -m^2_W) q^4}  = i \frac{g^{\mu' \nu'} }{4 (4\pi)^2  m^2_{H_6}} \frac{\ln y_W}{1-y_W} \,, 
 %
 \end{equation}
 the  $\Delta S=2$ effective Hamiltonian can be expressed as:
  \begin{equation}
  {\cal H}^{WH_6}_{\rm Box} 
   =  \frac{G^2_F (V^*_{ts} V_{td})^2 }{16\pi^2 }  m^2_W C_{WH_6} \bar s \gamma_\mu P_L d \bar s \gamma^\mu P_L d\,, \label{eq:Heff_WH6}
  \end{equation}
 where the Fierz transformation is applied and $y_{X} \equiv m_X^2/m_{H_6}^2$ for any particle $X$; 
 $G_F/\sqrt{2}=g^2/(8 m^2_W)$ is used, and the effective Wilson coefficient $C_{WH_6}$ at the $\mu=m_{H_6}$ scale  is given as: 
  \begin{equation}
  C_{W H_6} =- \frac{V^*_{j s} V_{i d} }{V^*_{ts} V_{td}} \frac{g^*_{j2} g_{i1}}{ g^2 V^*_{ts} V_{td}} \frac{4 y_W \ln y_W}{y_W -1} \,. \label{eq:CWH6}
  \end{equation}
 Unlike the case in the color-triplet diquark~\cite{Barr:1989fi,Chen:2018dfc}, the symmetric  color factor of $\delta^{\beta'}_\rho \delta^{\rho'}_\beta + \delta^{\rho'}_\rho \delta^{\beta'}_\beta$ from the color-sextet diquark does not lead to a cancellation. From Eq.~(\ref{eq:CWH6}), at first sight,   the box diagrams mediated by one $W$ and one ${\bf H}_6$  do not vanish in the limit of $m_{t}=  0$. However, it is found that Eq.~(\ref{eq:CWH6}) indeed vanishes. Using ${\bf g} =2 V^{u*}_L {\bf y} V^{d\dag}_L$ and $V=V^u_L V^{d\dag}_L$,  we can verify the result as follows:
  \begin{align}
  \sum_i V_{i d} g_{i 1} =2 \sum_{i} \left(V^{d*}_L V^{uT}_L\right)_{di} \left(V^{u*}_L {\bf y} V^{d\dag}_L \right)_{i1} =2\left(V^{d*}_L {\bf y} V^{d\dag}_L \right)_{11}=0\,,
  \end{align}
 where we use the quark-flavor indices to label the CKM matrix elements and use Arabic numerals  to show the Yukawa couplings. 
 The null result arises from the antisymmetric property of  $(V^{d*}_L {\bf y} V^{d\dag}_L)$; hence,  $C_{WH_6}=0$. 
 The result has nothing to do with the structure of ${\bf g}$ and is also suitable for $B$- and $D$-meson systems. Intriguingly, our conclusion is the same as that in~\cite{Barr:1989fi} using a different viewpoint. We should emphasize that it is ${\bf y}$, which was used in~\cite{Barr:1989fi}. It can be seen that if we use ${\bf y}$ instead of ${\bf g}$, $\sum_i V_{i d} y_{i 1}$ does not vanish.  For simplicity, we take $V^u_L=1$, $V=V^{d\dag}_L$, and ${\bf g}=2 {\bf y} V$ in the following analysis.  
 
 In the limit of $m_t=0$, the nonvanished contributions to $\Delta S=2$ are from the pure diquark box diagrams, and the effective Hamiltonian from the  ${\bf H}_{6}$  box diagrams can be simply written as:
 \begin{align}
  {\cal H}^{H_6 H_6}_{\rm Box}  &= \frac{G^2_F (V^*_{ts} V_{td})^2  }{16\pi^2 }  m^2_W C_{H_6 H_6} \bar s \gamma_\mu P_L d \bar s \gamma^\mu P_L d\,,\nonumber \\
C_{H_6 H_6} & = 6 y_W\left(  \frac{({\bf g^\dagger g})_{21}}{g^2 V^*_{ts} V_{td} }\right)^2 \,. \label{eq:Heff_H6}
  \end{align}
 Following the  hadronic matrix elements for $\Delta S=2$, which were obtained in~\cite{Buras:2001ra}, the matrix element of $K^0-\bar K^0$ mixing can be formulated as:
 \begin{align}
M^{K*}_{12} & = \langle \bar K^0 |   {\cal H}^{H_6 H_6}_{\rm Box} | K^0 \rangle  = \frac{G^2_F \left( V^*_{ts} V_{td} \right)^2}{48 \pi^2 } m^2_W m_K f^2_K  P^{VLL}_1  \nonumber  \\
& \times \left( \frac{\alpha_s(m_{H_6})}{\alpha_2(m_t)} \right)^{6/21} C_{H_6 H_6} \,,
 \end{align}
 where $f_K$ is the $K$-meson decay constant; $P^{VLL}_1 \approx 0.48$ is the nonperturbative QCD effects, and the factor $( \alpha_s(m_{H_6})/\alpha_2(m_t) )^{6/21}$ is the renormalization group (RG) evolution from the $m_{H_6}$ scale to the  $m_t$ scale. Since the Kaon indirect CP violation parameter  $\epsilon_K$ is associated with $V^*_{ts} V_{td}$ in the SM,  we scale the parameters $({\bf g^\dagger g})_{21}$ with the $(V^*_{ts} V_{td})^{-1}$ factor.
 Thus, the mass difference between $K_L$ and $K_S$ and $\epsilon_K$ can be obtained  as:
  \begin{align}
  \Delta M_{K} & = 2 Re M^K_{12}\,, ~\epsilon_K = \frac{\exp(i \pi/4) }{\sqrt{2} \Delta M_K}Im M^K_{12}\,.
  \end{align}
 The predicted $\Delta M_K$ in the SM is close to the experimental data; therefore, we use  $\Delta M^{\rm exp}_K$ instead of $\Delta M_K$, which appears in the denominator of $\epsilon_K$. In addition, the SM contribution to  $\epsilon_K$ is also consistent with the experimental data, so the room for new physics  actually is very limited. In our numerical analysis, we require that the new physics effects should satisfy~\cite{Buras:2015jaq}:
 \begin{align}
 |\Delta M^{\rm NP}_{K}| & \leq 0.2 \, \Delta M^{\rm exp}_K\,, \nonumber \\
 |\epsilon^{\rm NP}_{K}| & \leq 0.4 \times 10^{-3}\,. \label{eq:conditions}
 \end{align}

 Generally, ${\bf y}$ can carry two physical CP phases, for which a detailed discussion is given in the Appendix. If we require ${\bf y}$ to be a real matrix, from ${\bf g}=2 {\bf y} V$, it can be found that the origin of the CP violation  in the diquark model arises from the CKM matrix. Since it is not our purpose to show the generic CP phase effects of ${\bf y}$ in this study, we assume that   ${\bf y}$ is a real matrix in the following numerical analysis,  unless stated otherwise.  Taking the Wolfenstein's parametrization~\cite{Wolfenstein:1983yz} as an input, the CP phase of the CKM matrix appears in $V_{td}$ and $V_{ub}$, and the Kaon CP violation in the SM arises from $Im(V^*_{ts} V_{td})$; therefore,  when  the $\epsilon_K$ constraint is considered, we can simply focus on the $V_{td}$ related effects in $({\bf g^\dagger g})_{21}$. In addition, because  $\Delta M_K$ may cause a strict constraint on the free parameters, we should also pay attention to the effects, which arise from a large CP-conserving CKM factor.  Accordingly,  to study the $\Delta M_K$ and $\epsilon_K$ constraints, we decompose $({\bf g^\dagger g})_{21}$  as:
 \begin{align}
({\bf g^\dagger g})^{\rm CPC}_{21} & = 4 (V^\dagger {\bf y^\dagger y} V)^{\rm CPC}_{21} =4 V^*_{ks} V_{ud} ({\bf y^\dagger y})_{k1}+ 4 V^*_{ks}  V_{cd}({\bf y^\dagger y})_{k2}\,,  \nonumber \\
({\bf g^\dagger g})^{\rm CPV}_{21} & = 4 (V^\dagger {\bf y^\dagger y} V)^{\rm CPV}_{21} = 4 V^*_{ks} V_{td} ({\bf y^\dagger y})_{k 3}   \,,
\label{eq:gg21}
 \end{align}
where the sum of all $k$ flavors is implied, and $({\bf y^\dagger y})$ is a symmetric matrix.  It can be seen that if we use $({\bf y^\dagger y})_{ij}$ as the free parameters, there are nine components in $({\bf g^\dagger g})_{21}$; six terms are from $({\bf g^\dagger g})^{\rm CPC}_{21}$, and three terms are from $({\bf g^\dagger g})^{\rm CPV}_{21}$. We note that without any approximation, indeed $Im(V^*_{cs} V_{cd})=-Im(V^*_{ts} V_{td})$; however, since the associated new physics effects are small, we neglect the $Im(V^*_{cs} V_{cd})$ contributions. Thus, $({\bf g^\dagger g})^{\rm CPC}_{21}$ can be taken as a real parameter and $({\bf g^\dagger g})^{\rm CPV}_{21}$ carries the SM CP phase. 

From Eq.~(\ref{eq:gg21}), each $({\bf y^\dagger y})_{ij}$ is associated with a CKM factor $V^*_{is} V_{jd}$. It is known that the CKM matrix elements have a hierarchical structure; therefore, when the CKM factor is larger, the allowed $({\bf y^\dagger y})_{ij}$ is smaller. For clarity,  in terms of  the $\lambda$ parameter, we show the corresponding  CKM factor for the associated  $({\bf y^\dagger y})_{ij}$ 
in Table~\ref{tab:CKM}. It can be seen that although $({\bf y^\dagger y})_{13}$ and $({\bf y^\dagger y})_{33}$ can contribute to $\epsilon_K$, because the associated CKM factors are small,  their allowed values are much larger than the others. Nevertheless, due to ${\bf y}^T=-{\bf y}$, we can obtain:
 \begin{align}
 ({\bf y^\dagger y})_{13}& = y^*_{12} y_{23}\,, \nonumber \\
  ({\bf y^\dagger y})_{33} &=  |y_{13}|^2 + | y_{23}|^2 \geq 0\,,
 \end{align}
 where the same parameters also appear in $({\bf y^\dagger y})_{31}=y_{12} y^*_{23}$, $({\bf y^\dagger y})_{11}=|y_{12}|^2+|y_{13}|^2\geq 0$, and $({\bf y^\dagger y})_{22}=|y_{12}|^2+|y_{23}|^2 \geq 0$, which are strictly bounded by $\Delta M_K$. Hence, due to the antisymmetric nature of the ${\bf y}$ matrix, we cannot take all $g^*_{j2} g_{j1}$ as independent  parameters. 
 
 \begin{table}[hpbt]
\caption{ $\lambda$-parameter dependence of the CKM factor, which is associated with the parameter $({\bf y^\dagger y})_{ij}$. }
\label{tab:CKM}
\begin{tabular}{c|cccccccc} \hline \hline
 & $({\bf y^\dagger y})_{11}$~& ~$({\bf y^\dagger y})_{21}$~& ~$({\bf y^\dagger y})_{31}$~&~$({\bf y^\dagger y})_{12}$~  & ~$({\bf y^\dagger y})_{22}$ ~& ~$({\bf y^\dagger y})_{23(32)}$~ & ~$({\bf y^\dagger y})_{13}$~  &~$({\bf y^\dagger y})_{33}$ \\ \hline 
 CKM  &  $\lambda$ & $\lambda^0$  & $\lambda^2$ & $\lambda^2$ & $\lambda$ & $\lambda^3$  & $\lambda^4$ & $\lambda^5$ \\ \hline\hline
 \end{tabular}

\end{table}

\begin{figure}[phtb]
\includegraphics[scale=0.60]{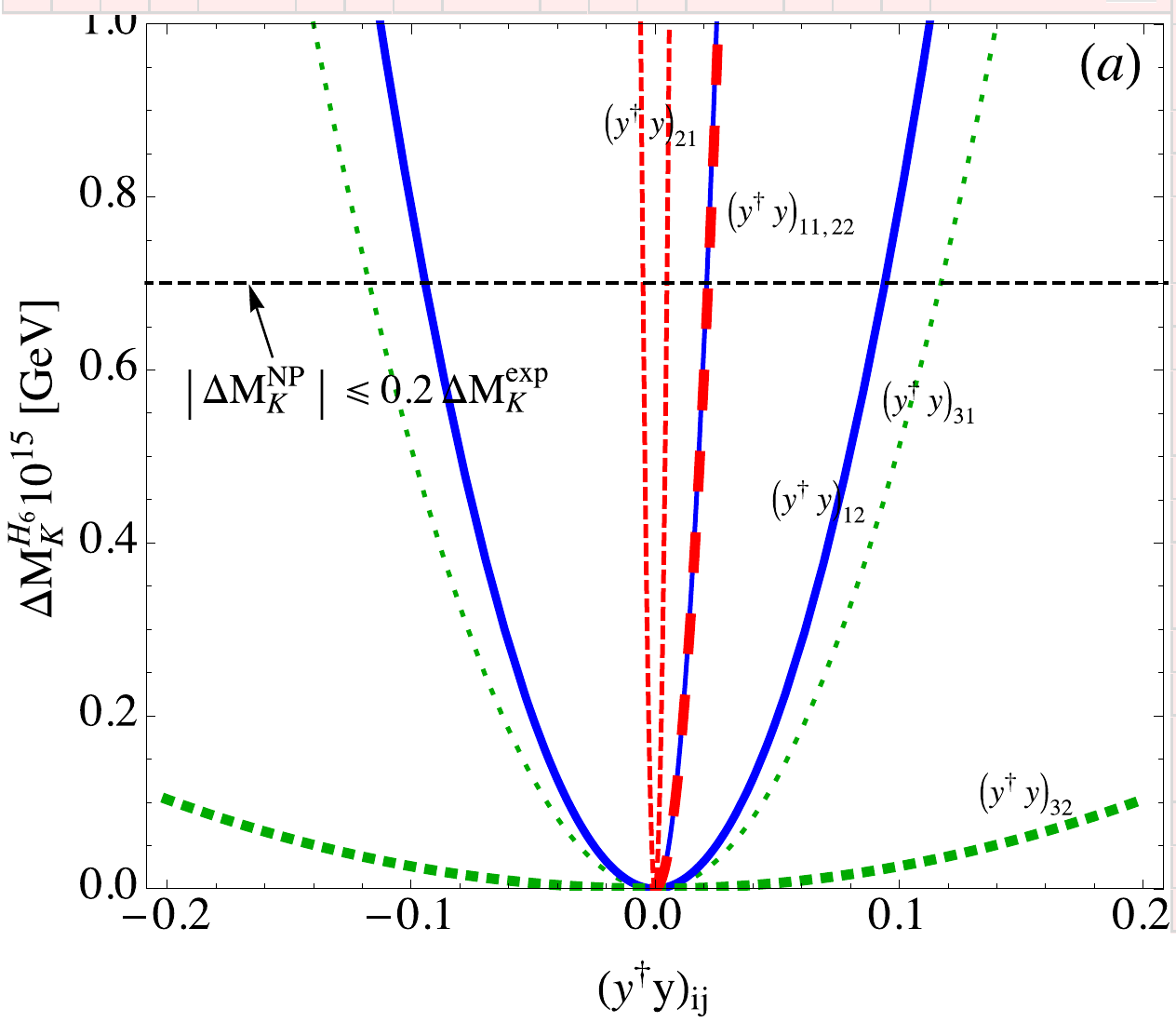}
\includegraphics[scale=0.60]{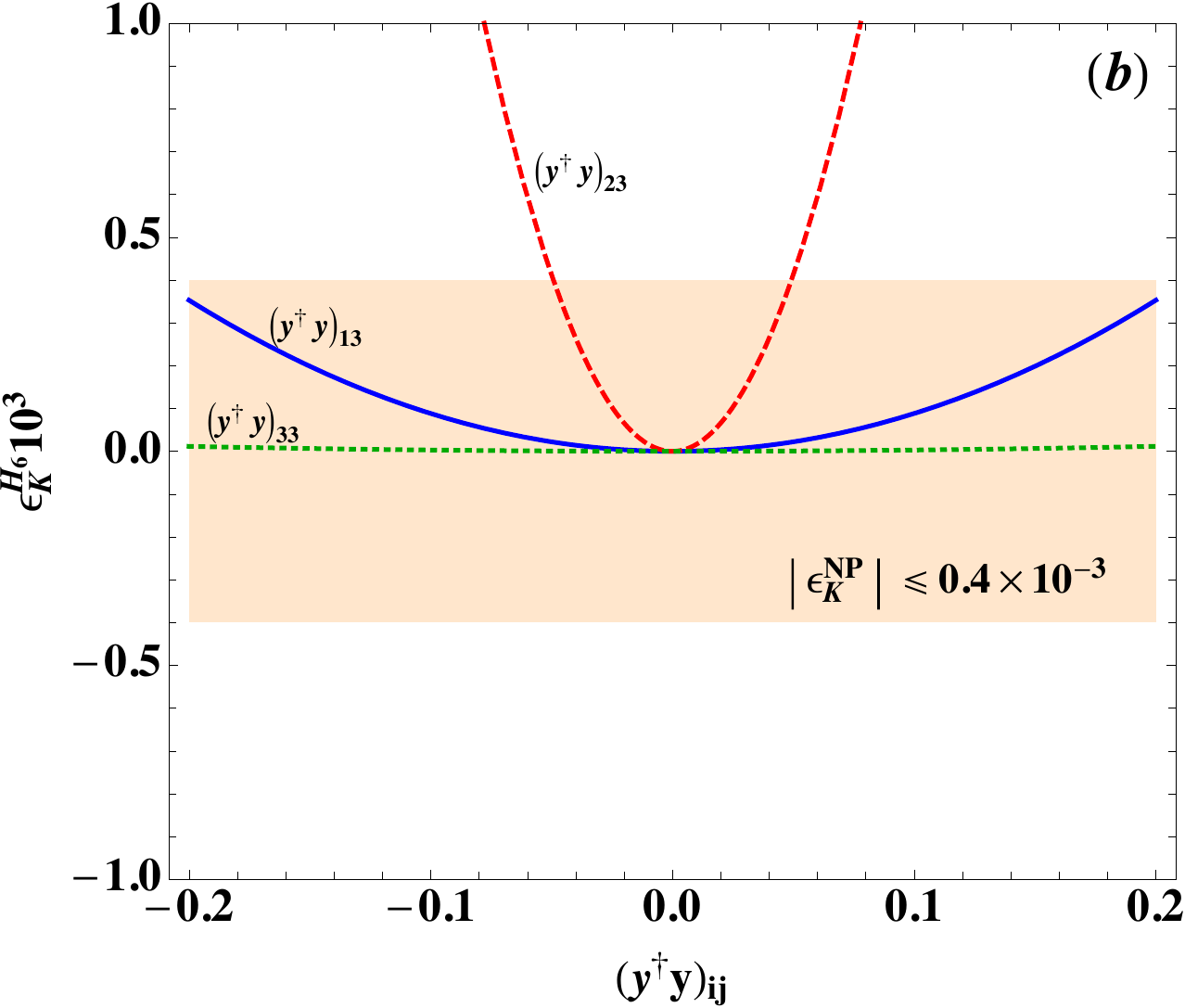}
 \caption{ (a) $\Delta M^{\rm H_6}_{K}$ (in units of $10^{-15}$) and (b) $\epsilon^{\rm H_6}_{K}$ (in units of $10^{-3})$ as a function of  $({\bf y^\dagger y})_{ij}$, where $m_{H_6}=1.5$ TeV is used, where the horizontal dashed line in (a) is the upper limit of $\Delta M^{\rm NP}_{K} \leq 0.2 \Delta M^{\rm exp}_{K}$, and the shaded area in (b) denotes $|\epsilon^{\rm NP}_K|\leq 0.4 \times 10^{-3}$.  }
\label{fig:yy_CP_ab}
\end{figure}

To determine the magnitude of each involved $({\bf y^\dagger y})_{ij}$ when the $\Delta M_K$ and $\epsilon_K$ constraints are satisfied,  we take all the six terms of $({\bf g^\dagger g})^{\rm CPC}_{21}$ and the three terms of  $({\bf g^\dagger g})^{\rm CPV}_{21}$ in Eq.~(\ref{eq:gg21}) as independent effects. At the moment, we do not consider the possible cancellations among different parameters. 
 According to the recent CMS measurement~\cite{Sirunyan:2018xlo}, the upper limit on $\sigma A {\cal B}$ for the diquark resonance  decaying to quark-quark at $m_{H_3}\approx 1.5$ TeV is given by $\sim 10$ pb, where the electromagnetic coupling is applied, and $\sigma$, $A$, and ${\cal B}$ denote the production cross section, acceptance, and branching fraction, respectively. Using  the calculating results in~\cite{Han:2009ya}, we obtain $\sigma\sim 1.16$ pb  when $g_{11}\approx 2 y_{12}V_{21}\approx 0.176$ is used, where the corresponding CMS limit is $\sigma \sim 3.09$ pb. 
 In our following numerical estimates, we  fix $m_{H_6}=1.5$ TeV.
Thus, each diquark contribution to $\Delta M_K$ from $({\bf g^\dagger g})^{\rm CPC}_{21}$  is shown as a function of $({\bf y^\dagger y})_{ij}$ in Fig.~\ref{fig:yy_CP_ab}(a). The solid, dashed, and dotted lines denote the contributions from $({\bf y^\dagger y})_{11}$, $({\bf y^\dagger y})_{21}$, and $({\bf y^\dagger y})_{31}$,  whereas the thick solid,  dashed, and dotted are the contributions from $({\bf y^\dagger y})_{12}$, $({\bf y^\dagger y})_{22}$, and $({\bf y^\dagger y})_{32}$, respectively. Due to the same CKM factor, the lines of $({\bf y^\dagger y})_{11}$ (solid) and $({\bf y^\dagger y})_{22}$ (thick dashed) overlap.  The results with the $\epsilon_K$ constraint are shown in Fig.~\ref{fig:yy_CP_ab}(b), where the solid, dashed, and dotted lines denote the effects from $({\bf y^\dagger y})_{13}$, $({\bf y^\dagger y})_{23}$, and $({\bf y^\dagger y})_{33}$, respectively. Due to $({\bf y^\dagger y})_{ij}=({\bf y^\dagger y})_{ji}$, we see that the constraint on $({\bf y^\dagger y})_{13}$ from $\Delta M_K$ is stricter than that from $\epsilon_K$. As mentioned earlier, $({\bf y^\dagger y})_{33}=|y_{12}|^2 + |y_{13}|^2$ is correlated to $({\bf y^\dagger y})_{11}=|y_{13}|^2 + |y_{23}|^2$ and $({\bf y^\dagger y})_{22}=|y_{12}|^2 + |y_{23}|^2$, so, although large ranges of $y_{13}$ and $y_{23}$ in $({\bf y^\dagger y})_{33}$ are allowed under the $\epsilon_K$ constraint, they indeed have been  bounded by $({\bf y^\dagger y})_{11, 22}$ through $\Delta M_K$. Hence, with the exception of $({\bf y^\dagger y})_{23}$, it is  found that   the constraint from $\Delta M_K$ is more serious than that from $\epsilon_K$, and the resulting upper values of  $|({\bf y^\dagger y})_{ij}|$ are generally below 0.12.  

 \subsection{ $m_t \neq 0$}
 
We have discussed that in the limit of $m_t=0$,   the $W$-${\bf H_6}$ box diagrams for $\Delta S=2$ vanish. 
 It is of interest to see if the box diagrams can have a sizable effect when  $m_t(\bar m_t)\approx 165$ GeV is used. In addition to the $W$-mediated  box diagrams, in the 't Hooft-Feynman gauge, we now have to include the charged-Goldstone-boson contribution. Since the Goldstone-boson Yukawa coupling to a quark is proportional to the quark mass, we only need to calculate the top-quark contribution.  Hence, the effective Hamiltonian for $\Delta S=2$ can be written as:
 \begin{align}
  {\cal H}^{(W+G)H_6}_{\rm Box} &= {\cal H}^{WH_6}_{\rm Box} + {\cal H}^{GH_6}_{\rm Box}  \nonumber \\
  & = \frac{G^2_F (V^*_{ts} V_{td})^2}{16\pi^2} m^2_W \left(C_{W H_6} + C_{GH_6} \right)\bar s \gamma_\mu P_L d\, \bar s \gamma^\mu P_L d \,, \nonumber \\
 C_{WH_6} & = - 8 y_W h_{ij} \frac{V^*_{js} V_{id}}{V^*_{ts} V_{td}}I_{WH_6}(y_W, y_{u_i},y_{u_j})\,, \nonumber \\
 C_{G H_6} & =  - 4 y^2_t h_{33} I_{GH_6}(y_W,y_t) \,, \label{eq:CWH6_mt}
 %
 \end{align}
where $h_{ij}=g^*_{j2} g_{i1}/(g^2 V^*_{ts} V_{td})$, the sum of $i,j$ is implied, and  the loop functions are shown as:
  \begin{align}
  I_{WH_6}(y_W, y_{u_i},y_{u_j}) & = \int^1_0 dx_1 \int^{x_1}_{0} dx_2 \frac{x_2}{1+ (y_W-1)x_1 + (y_{u_i}-y_W) x_2 + (y_{u_j}-y_{u_i})x_3}\,, \nonumber \\
  I_{GH_6}(y_W, y_t) & =  \int^1_0 dx_1 \int^{x_1}_{0} dx_2 \frac{x_2}{(1+ (y_W-1)x_1 + (y_t-y_W) x_2)^2} \,.  %
  \end{align}
  
  Because $C_{WH_6}=0$ when $m_{u,c,t}=0$,  it is expected that  $C_{WH_6}$ with $m_t\neq 0$ can be expressed in terms of the difference between the $m_t=0$ and $m_t \neq 0$ cases. From the identity of $( V^T {\bf g})_{ii}=0$, we can see that $\sum_{k'} V_{k'd} g_{k'1}=-V_{td} g_{31}$, and $\sum_{k'} V_{k's} g_{k'2}=-V_{ts} g_{32}$, where $k'$ denotes the flavors of the first two generations. Accordingly, $C_{WH_6}$ in Eq.~(\ref{eq:CWH6_mt}) can be formulated as:
   \begin{equation}
   C_{WH_6}  = - 8 y_W  h_{33} \left( I^0_{WH_6}(y_W) + I^{tt}_{WH_6}(y_W, y_t) -2 I^{t}_{WH_6}(y_W, y_t) \right)\,, 
\end{equation}
with
\begin{align}
   I^0_{WH_6} (y_W)&=I_{WH_6}(y_W,0,0)\,,  \nonumber \\
   I^t_{WH_6} (y_W, y_t))& = I_{WH_6}(y_W,y_t,0) = I_{WH_6}(y_W, 0, y_t)\,, \nonumber \\
   I^{tt}_{WH_6} (y_W, y_t) & = I_{WH_6}(y_W, y_t, y_t)\,.
   \end{align}
It can be seen that in the case of  $m_t \neq 0$, the nonvanished $C_{WH_6}$ is related to $g^*_{32} g_{31}$, and  the relationship to $({\bf y^\dagger y})_{ij}$ can be obtained as:
\begin{align}
g^*_{32} g_{31} & = 4({\bf y}V)^*_{32} ({\bf y} V)_{31}  \nonumber \\
& \approx  4\left( (|y_{31}|^2 - |y_{32}|^2) \lambda + y^*_{31} y_{32} \lambda^2 + y^*_{32} y_{31} \right) \nonumber \\
& \approx 4 ({\bf y^\dagger y})_{21} + 4 [({\bf y^\dagger y})_{11}-({\bf y^\dagger y})_{22}]\lambda \,, \label{eq:g32g31}
\end{align}
where  $V_{ud,cs}\approx 1$ and $V_{cd}\approx -V_{ud} \approx -\lambda$ are used, and the $\lambda^2$ term has been neglected.  If we drop the $\lambda$-related subleading effects, we can obtain  $g^*_{32} g_{31}\approx 4 ({\bf y^\dagger y})_{21}$.

\begin{figure}[phtb]
\includegraphics[scale=0.65]{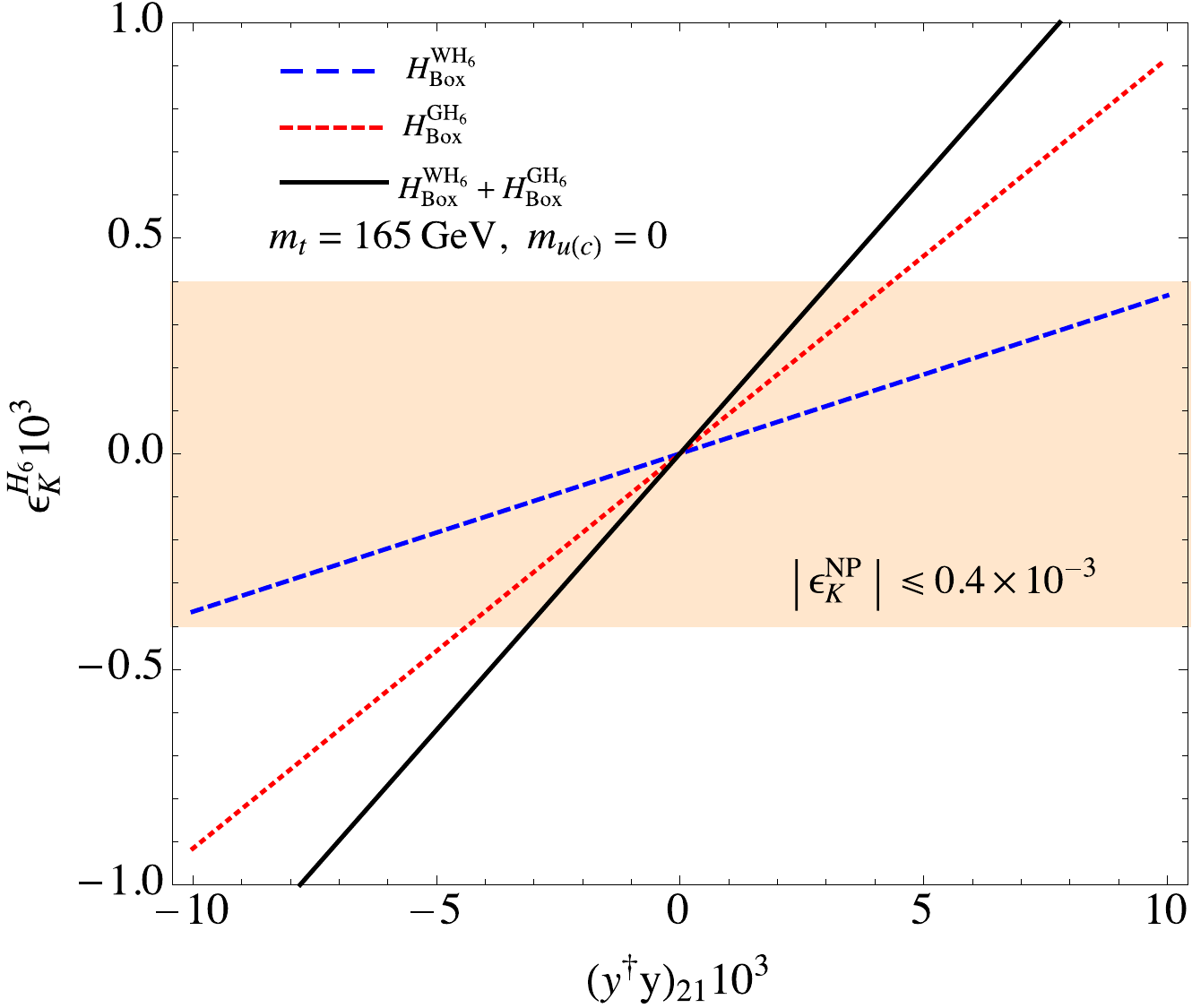}
 \caption{ $W$-${\bf H}_6$ (dashed) and $G$-${\bf H}_6$ (dotted)  box diagrams contributing to $\epsilon_K$ (in units of $10^{-3}$), where the solid line denotes the results of ${\cal H}^{(W+G)H_6}_{\rm Box}$, and $m_t =165$ GeV and $m_{H_6}=1.5$ TeV are used.  }
\label{fig:Box_mt}
\end{figure}

  Using the results in~\cite{Buras:2001ra}, the $K^0-\bar K^0$ mixing matrix element in the diquark model can be expressed as:
  \begin{align}
  \langle \bar K^0 | {\cal H}^{(W+G) H_6}_{\rm Box} | K^0 \rangle & = \frac{G^2_F \left(V^*_{ts} V_{td} \right)^2 }{48 \pi^2} m^2_W m_K  f^2_K P^{VLL}_1 \left( \frac{\alpha_s(m_{H_6})}{\alpha_2(m_t)} \right)^{6/21}  \left(C_{W H_6} + C_{GH_6}  \right) \,, \\
C_{W H_6} + C_{GH_6} & \approx  -16  \frac{({\bf y^\dagger y})_{21} }{g^2 V^*_{ts} V_{td}} \left[ 2 y_W \left( I^0_{WH_6}(y_W) + I^{tt}_{WH_6}(y_W, y_t) -2 I^{t}_{WH_6}(y_W, y_t) \right)\right.\nonumber \\
& \left. +  y^2_t I_{GH_6}(y_W,y_t) \right]\,.
  \end{align}
  With $m_{H_6}=1.5$ TeV, the values of the loop functions are estimated to be $I^0_{WH_6}\approx 2.93$, $I^t_{WH_6}\approx 2.02$, $I^{tt}_{WH_6}\approx 1.67$, and  $I_{GH_6}\approx 56.51$. Although $I_{GH_6}$ is much larger than the others, when including the $y^2_t$ factor, the charged-Goldstone-boson contribution is close to that  derived from the $W$-${\bf H}_6$ box diagrams. To see the influence of the heavy top-qaurk, we plot $\epsilon^{\rm NP}_K$ (in units of $10^{-3}$) as a function of $({\bf y^\dagger y})_{21}$ ( in units of $10^{-3})$ in Fig.~\ref{fig:Box_mt}, where  the dashed and dotted lines denote the results from ${\cal H}^{WH_6}_{\rm Box}$ and ${\cal H}^{GH_6}_{\rm Box}$, respectively, and the solid line is the combined results. A comparison with the results in Fig.~\ref{fig:yy_CP_ab}(a) shows that the  constraint on $({\bf y^\dagger y})_{21}$ from ${\cal H}^{(W+G)H_6}_{\rm Box}$ is as strict as that from ${\cal H}^{H_6H_6}_{\rm Box}$ through $\Delta M_K$. That is, $W(G)-{\bf H}_6$ box diagrams indeed are important in the color-sextet diquark model when the heavy top-quark mass effects are included. We noted that the pure ${\bf H}_6$-mediated box diagrams are insensitive to the $m_t$ effects. We checked that the difference between  $m_t=0$ and $m_t=165$ GeV is only around $5\%$; therefore, we did not  reanalyze the ${\cal H}^{H_6 H_6}_{\rm Box}$ contributions with $m_t \neq 0$. 
 
\subsection{ Right-handed color-sextet diquark \label{sec:right-handed}}

 In addition to the left-handed couplings, ${\bf H}_6$ can also couple to the right-handed quarks. In order to understand the influence of  the right-handed couplings, we investigate the contributions to the $\Delta S=2$ processes in this section. The relevant Yukawa couplings can be written as:
\begin{equation}
-{\cal L}_Y \supset \overline{d^C_{Ri}} g^R_{ji} {\bf H^\dagger}_{6}  u_{Rj} + \bar u_{Rj} g^{R*}_{ji} {\bf H}_6 d^C_{Ri}\,.
\end{equation}
Unlike the left-handed diquark, which the Yukawa couplings are antisymmetric in flavor space, $g^R_{ij}$ have no particular symmetry in flavor indices.  Since we do not have any information about the right-handed flavor mixing matrices, which are used to diagonalize the mass matrices, $g^R_{ij}$ can be taken as the general Yukawa couplings when the right-handed up- and down-type quarks are the physical eigenstates. Thus, from the $W(G)$-${\bf H}_6$ and ${\bf H}_6$-${\bf H}_6$ box diagrams,  the effective Hamiltonian for $\Delta S=2$ at the $\mu=m_{H_6}$ scale can be  obtained as:
\begin{align}
 {\cal H}^{H_6}_{\Delta S=2} & = \frac{G^2_F (V^*_{ts} V_{td})^2 }{16 \pi^2 } m^2_W \left[ (C_{WH_6} + C_{GH_6} + C_{H_6 H_6}) Q^{VLL}_1 + C^{VRR}_{H_6 H_6} Q^{VRR}_1    \right. \nonumber \\
 & \left. + \left(C^{LR}_{1 H_6 H_6} +C^{LR}_{1WH_6} + C^{LR}_{1GH_6} \right) Q^{LR}_{1}    + \left(C^{LR}_{2 H_6 H_6} +C^{LR}_{2WH_6} + C^{LR}_{2GH_6} \right) Q^{LR}_2 \right] \,,
  \label{eq:H_H6_2}
 \end{align}
 where the left- and right-handed couplings are included; the effective operators are defined as:
 \begin{align}
 Q^{VLL}_	1 & = (\bar s \gamma_\mu P_L d) (\bar s \gamma^\mu P_L d)\,, ~ Q^{VRR}_1 = (\bar s \gamma_\mu P_R d) (\bar s \gamma^\mu P_R d)\,, \nonumber \\
 Q^{LR}_{1} & =  (\bar s \gamma_\mu P_L d) (\bar s \gamma^\mu P_R d)\,, ~  Q^{LRT}_{2}= (\bar s  P_L d) (\bar s  P_R d)\,;
 \end{align}
   the Wilson coefficients $C_{H_6 H_6}$ and  $C_{W(G) H_6}$ can be found in Eqs.~(\ref{eq:Heff_H6}) and (\ref{eq:CWH6_mt}),  and  the other new Wilson coefficients due to the right-handed couplings are given as:
  \begin{align}
  C^{VRR}_{H_6 H_6} & = 6 y_W \left( \frac{ ({\bf g^{R\dagger} g^R})_{21} }{g^2 V^*_{ts} V_{td}}\right)^2\,, ~ C^{LR}_{1H_6 H_6} = 2 y_W \frac{{(\bf g^\dagger g})_{21} }{g^2 V^*_{ts} V_{td} } \frac{{(\bf g^{R\dagger} g^R})_{21} }{g^2 V^*_{ts} V_{td} } \,, \nonumber \\
  C^{LR}_{1WH_6} & = -4  \frac{g^{R*}_{32} g^R_{31}}{g^2 V^*_{ts} V_{td} }   y_Wy_t  I_{G H_6}(y_W, y_t) \,, ~ C^{LR}_{1GH_6}  = -4  \frac{g^{R*}_{32} g^R_{31}}{g^2 V^*_{ts} V_{td} }  y_t  I_{W H_6}(y_W, y_t, y_t) \,, \nonumber \\
  C^{LR}_{2H_6 H_6} & = -20 y_W  \frac{{(\bf g^\dagger g})_{21} }{g^2 V^*_{ts} V_{td} } \frac{{(\bf g^{R\dagger} g^R})_{21} }{g^2 V^*_{ts} V_{td} }\,, ~ C^{LR}_{2WH_6} = - C^{LR}_{1WH_6}\,, ~ C^{LR}_{2GH_6} = - C^{LR}_{1GH_6}\,. \label{eq:R_WCs}
  \end{align}

 Similar to the case in the left-handed diquark, the ${\bf H}_6$-${\bf H}_6$ diagrams are insensitive to $m_t$; therefore, we use $m_{t}=0$ in the ${\bf H}_6$-${\bf H}_6$ box diagrams. We note that $C^{LR}_{1(2) WH_6}$ and $C^{LR}_{1(2) GH_6}$ are from the $W(G)$-${\bf H}_6$ box diagrams. Because their effects are proportional to quark masses, which arise from the fermion propagators, the light quark contributions can be neglected; therefore,  they only depend on $g^{R*}_{32} g^R_{31}$. Moreover, due to the $W(G)$-${\bf H}_6$ diagrams being associated with the CKM factor $V^*_{ts} V_{td}$, even $g^{R*}_{32} g^R_{31}$ is a real parameter,  $C^{LR}_{1(2) WH_6}$ and $C^{LR}_{1(2) GH_6}$ can contribute to $\epsilon_K$ through the SM CP violating phase.  Similarly, the SM CP-phase hidden in $({\bf g}^\dagger {\bf g})_{21}$ of  $C^{LR}_{1(2)H_6 H_6}$, which are from the ${\bf H}_6$-${\bf H}_6$ box diagrams, can also contribute to $\epsilon_K$. 
   
 Using the matrix elements obtained in~\cite{Buras:2001ra},  the $K^0-\bar K^0$ mixing matrix element  can  then be expressed as:
  \begin{align}
  \langle \bar K^0 | {\cal H}_{\rm Box} | K^0 \rangle & = 
  \frac{G^2_F \left(V^*_{ts} V_{td} \right)^2 }{48 \pi^2} m^2_W m_K  f^2_K \left[ P^{VLL}_1  \left( C^{VLL}_1(m_t) + C^{VRR}_1 (m_t) \right)  \right. \nonumber \\
  &    \left.  + P^{LR}_1C^{LR}_1(m_t)  + P^{LR}_{2} C^{LR}_2(m_t) \right] \,,
  \end{align}
  where  the $\{C_{i}\}$ coefficients are defined as:
   \begin{align}
   C^{VLL}_{1}(m_t) &= \eta^{6/21}\left( C_{WH_6} + C_{G H_6} + C_{H_6 H_6} \right)\,, ~C^{VRR}_{1} (m_t)=  \eta^{6/21} C^{VRR}_{H_6 H_6} \,,  \nonumber \\
   C^{LR}_{1} (m_t)& =   \eta^{3/21}  \left( C^{LR}_{1H_6 H_6} + C^{LR}_{1 WH_6} + C^{LR}_{1GH_6}\right)\nonumber \\
   C^{LR}_{2} (m_t)& =  \frac{2}{3} \left( \eta^{3/21} - \eta^{-24/21} \right) \left( C^{LR}_{1H_6 H_6} + C^{LR}_{1 WH_6} + C^{LR}_{1GH_6}\right) \nonumber \\
   & +\eta^{-24/21} \left( C^{LR}_{2H_6 H_6} + C^{LR}_{2 WH_6} + C^{LR}_{2GH_6}\right) \,,
   \end{align}
 with $\eta=\alpha_s(m_{H_6})/\alpha_s(m_t)$,  and the nonperturbative hadronic effects parametrized by $\{P_{i}\}$ are given as:
   \begin{equation}
   P^{VLL}_1 = 0.48\,, ~ P^{LR}_1 = -36.1\,, ~ P^{LR}_2 = 59.3\,. 
   \end{equation}
   It can be seen that $|P^{LR}_{1(2)}| \gg P^{VLL}_1$, where the large $|P^{LR}_{1,2}|$ values arise from the  enhancement factor $m^2_K/(m_s(\mu) + m_s(\mu))^2$. 
  
\begin{figure}[phtb]
\includegraphics[scale=0.5]{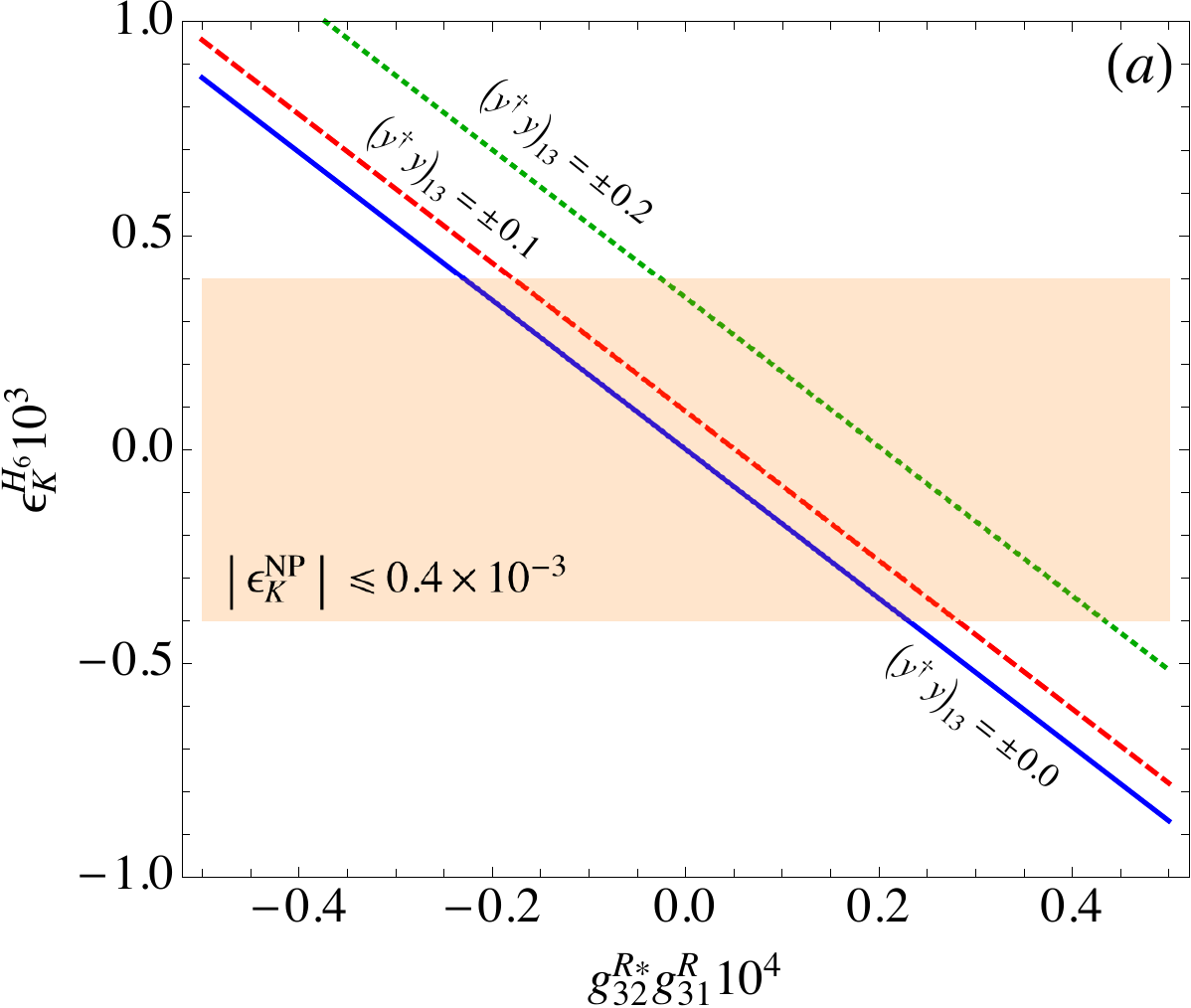}
\includegraphics[scale=0.5]{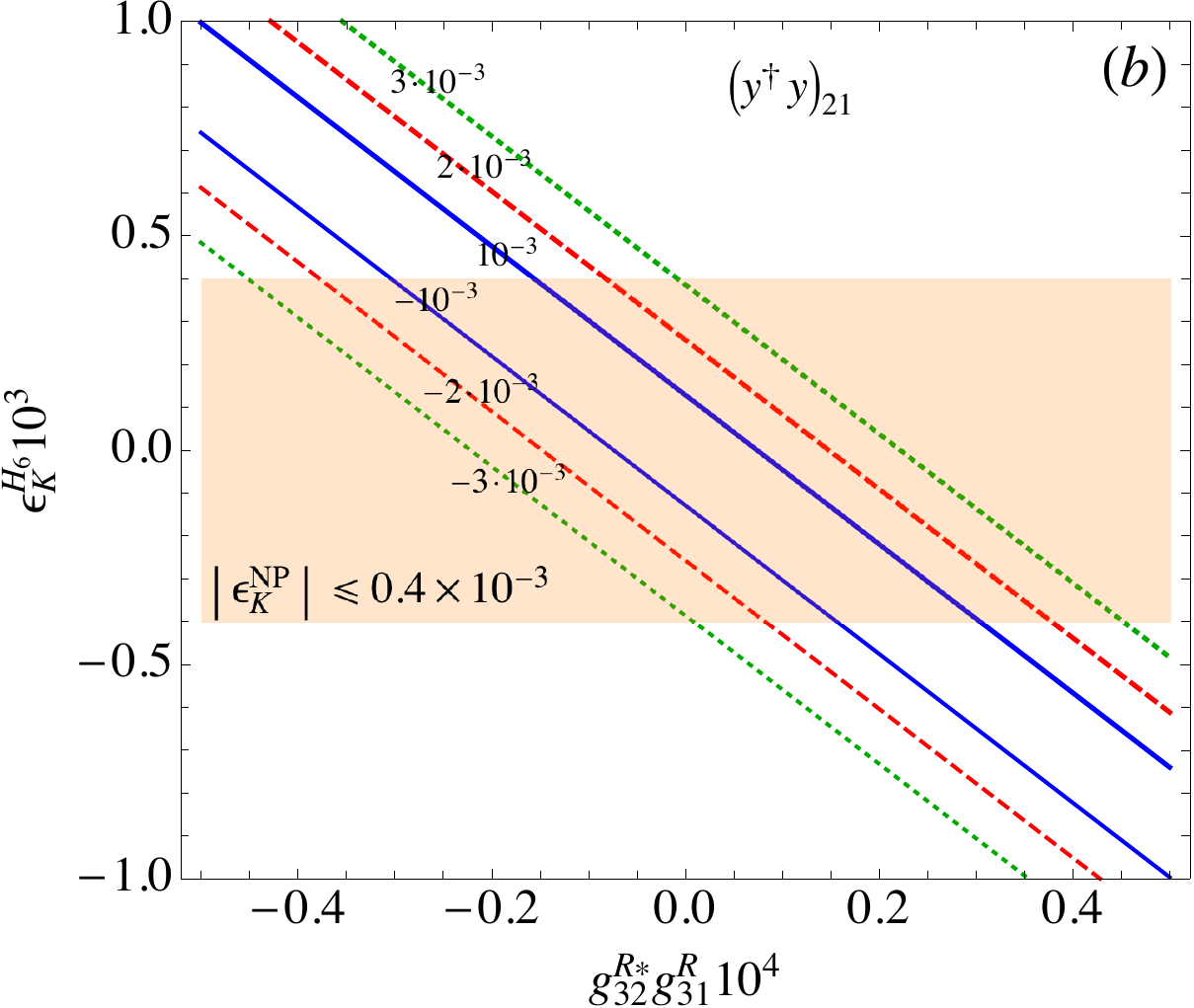}
\includegraphics[scale=0.5]{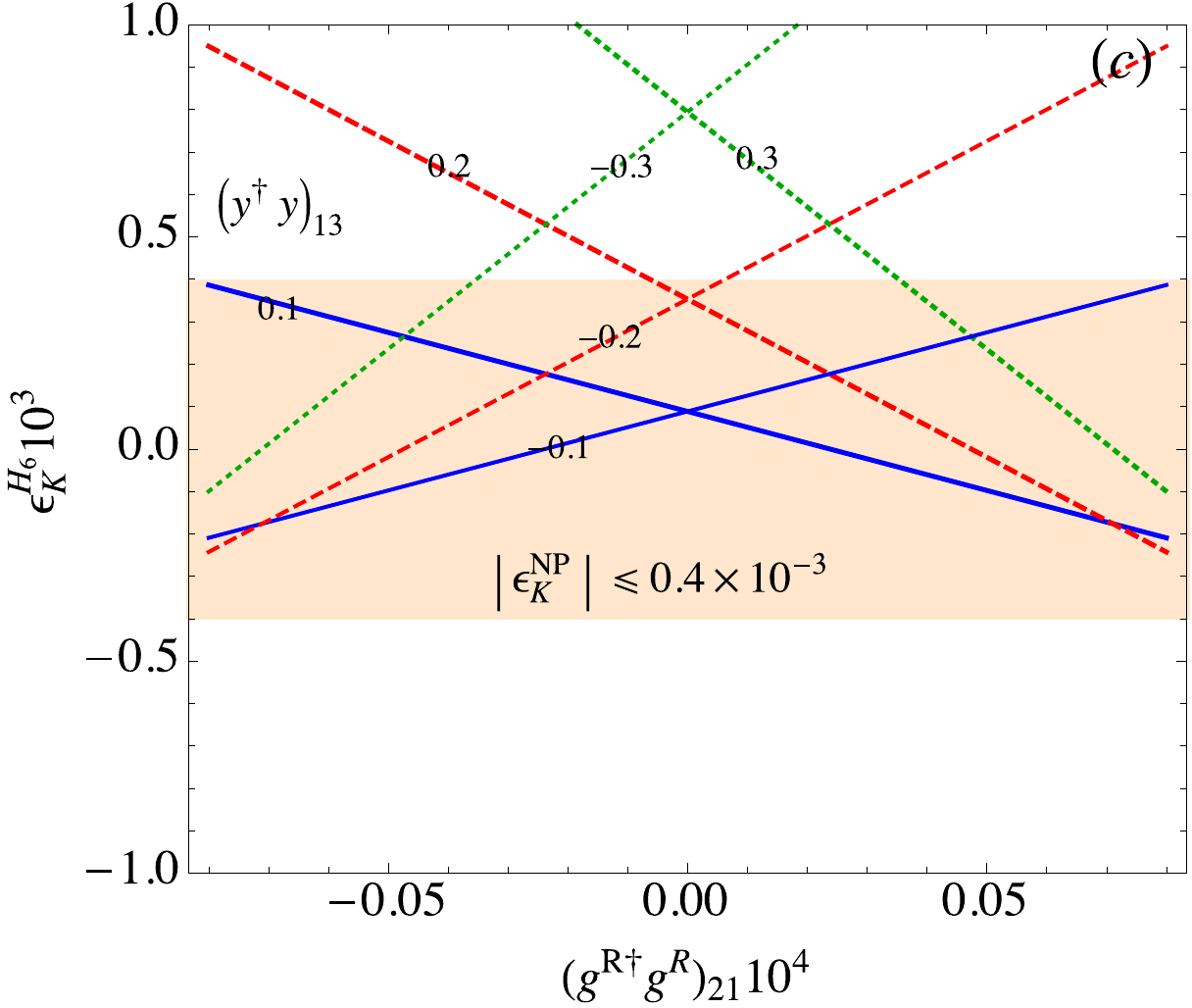}
\includegraphics[scale=0.5]{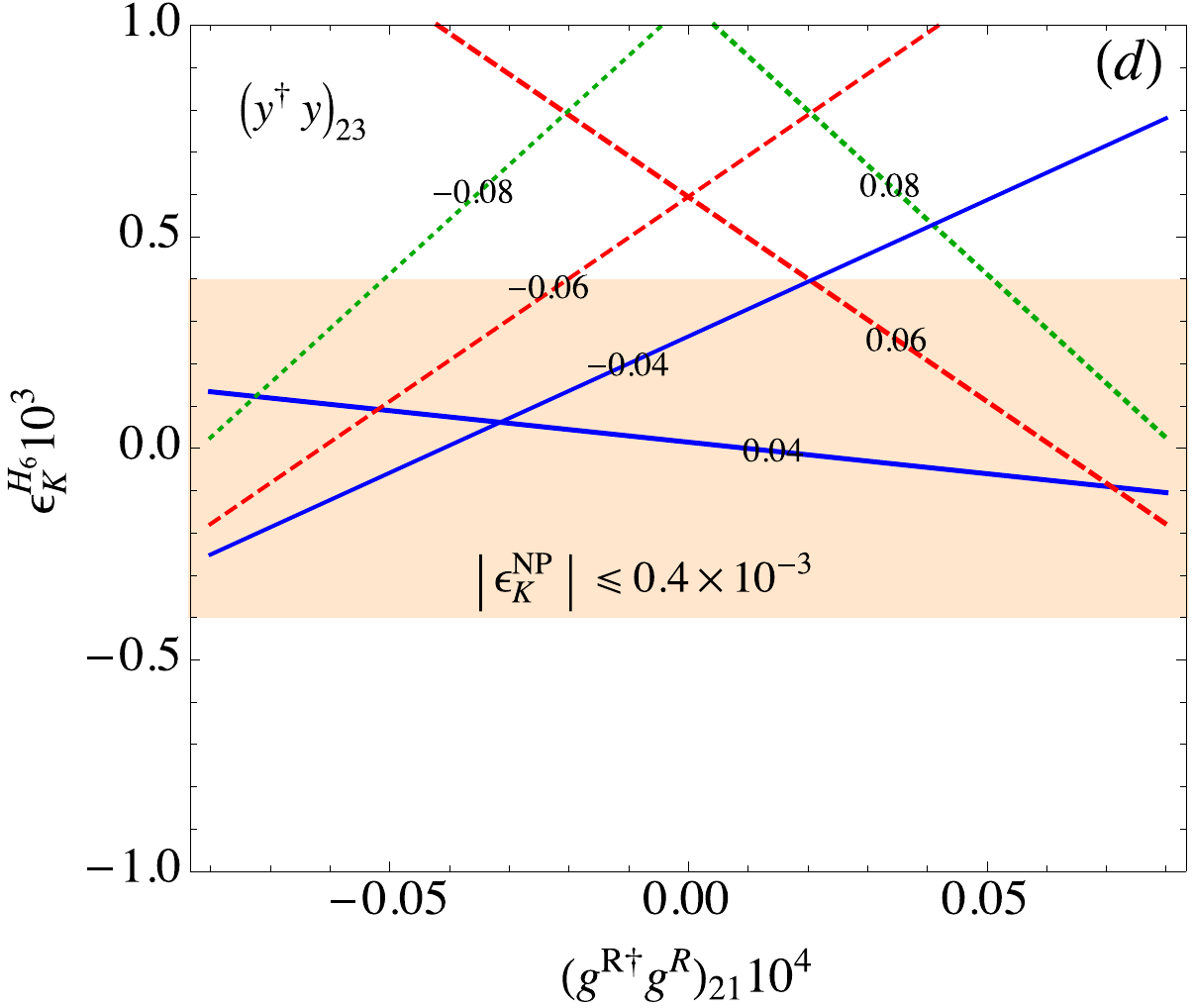}
 \caption{  $\epsilon^{H_6}_K$ (in units of $10^{-3}$) as a function of $g^{R*}_{32} g^R_{31}$ shown in plots (a) and (b) and as a function of $({\bf g}^{R\dagger} {\bf g}^R)_{21}$ shown  in plots (c) and (d), where the values on the lines denote the taken benchmarks of $({\bf y}^\dagger {\bf y})_{ij}$.}
\label{fig:R_CP-odd}
\end{figure}

  Since the CKM matrix elements do not appear in $g^{R*}_{32} g^R_{31}$ and $({\bf g}^{R\dagger} {\bf g}^R)_{21}$, we assume $g^R_{ij}$ to be real parameters. Nevertheless, the $W(G)$-${\bf H}_6$ and ${\bf H}_6$-${\bf H}_6$ diagrams are still associated with the $V_{td}$ CKM matrix element; thus,  it can be expected that   $\epsilon_K$  gives a  stricter constraint on $g^R_{ij}$ comparing with  $\Delta M_K$.  In order to illustrate the constraints on the right-handed couplings, we show $\epsilon^{H_6}_K$ (in units of $10^{-3}$) as a function of $g^{R*}_{32}g^R_{31}$ in Fig.~\ref{fig:R_CP-odd}(a) and (b), where $({\bf g}^{R\dagger} {\bf g}^R)_{21} =0$ is used for simplicity, and  some benchmark values of $({\bf y}^\dagger {\bf y})_{13}$ and $({\bf y}^\dagger {\bf y})_{21}$, which  satisfy the limits shown in Figs.~\ref{fig:yy_CP_ab}(b) and \ref{fig:Box_mt}, are taken. We note that in order to see the influence of each parameter, as analyzed earlier, when we use one of $({\bf y}^\dagger {\bf y})_{ij}$ in the numerical analysis, the others are taken to be zero. From the results, it can be found that $|g^{R*}_{32} g^R_{31}| < 0.5\times 10^{-4}$.  
  
 In order to singly analyze the $({\bf g}^{R\dagger} {\bf g}^R)_{21}$ constraint, we set $g^{R*}_{32} g^R_{31}=0$ although it is included in $({\bf g}^{R\dagger} {\bf g}^R)_{21}$. From Eq.~(\ref{eq:R_WCs}),  it can be seen that with the exception of $C^{VRR}_{H_6 H_6}$,  the remaining Wilson coefficients $C^{LR}_{1H_6 H_6}$ and $C^{LR}_{2H_6 H_6}$ are related to the product of $({\bf g}^{R\dagger} {\bf g}^R)_{21}$ and $({\bf g}^\dagger{\bf g})_{21}$.  If we first neglect the $({\bf g}^\dagger{\bf g})_{21}$ effect, the $({\bf g}^{R\dagger} {\bf g}^R)_{21}$ in $C^{VRR}_{H_6 H_6}$ can only contribute to $\Delta M_K$, and it can be found that  the bound is  $|({\bf g}^{R\dagger} {\bf g}^R)_{21}|<0.02$.   To investigate the $\epsilon_K$ constraint, from Eq.~(\ref{eq:gg21}), we can focus on the $({\bf g}^\dagger{\bf g})^{\rm CPV}_{21}$ parts. Accordingly, including $({\bf g}^\dagger{\bf g})_{21}$ effect, we show  $\epsilon^{H_6}_K$ as a function of $({\bf g}^{R\dagger} {\bf g}^R)_{21}$ in Fig.~\ref{fig:R_CP-odd}(c) and (d) with some benchmark values of $({\bf y}^\dagger{\bf y})_{13}$ and $({\bf y}^\dagger{\bf y})_{23}$, respectively, where the taken values satisfy the results shown in Fig.~\ref{fig:yy_CP_ab}(b). From the results, it can be seen that the allowed $({\bf g}^{R\dagger} {\bf g}^R)_{21}$ region combined  with the $({\bf y}^\dagger{\bf y})_{ij}$ effects is much smaller than   $g^{R*}_{32} g^R_{31}$ when $({\bf g}^{R\dagger} {\bf g}^R)_{21}=0$ is assumed. 

Since the allowed $g^{R*}_{32} g^R_{31}$ and $({\bf g}^{R\dagger} {\bf g}^R)_{21}$ regions are much smaller than the left-handed diqurak couplings $({\bf y}^\dagger {\bf y})_{ij}$, in order to emphasize the peculiar antisymmetryic property of ${\bf y}$, in the following analysis, we assume $g^R_{ij}$ to be small and neglect their effects.  
  
\subsection{Parameter scan with the $\Delta S=2$ constraints}

\begin{figure}[phtb]
\includegraphics[scale=0.60]{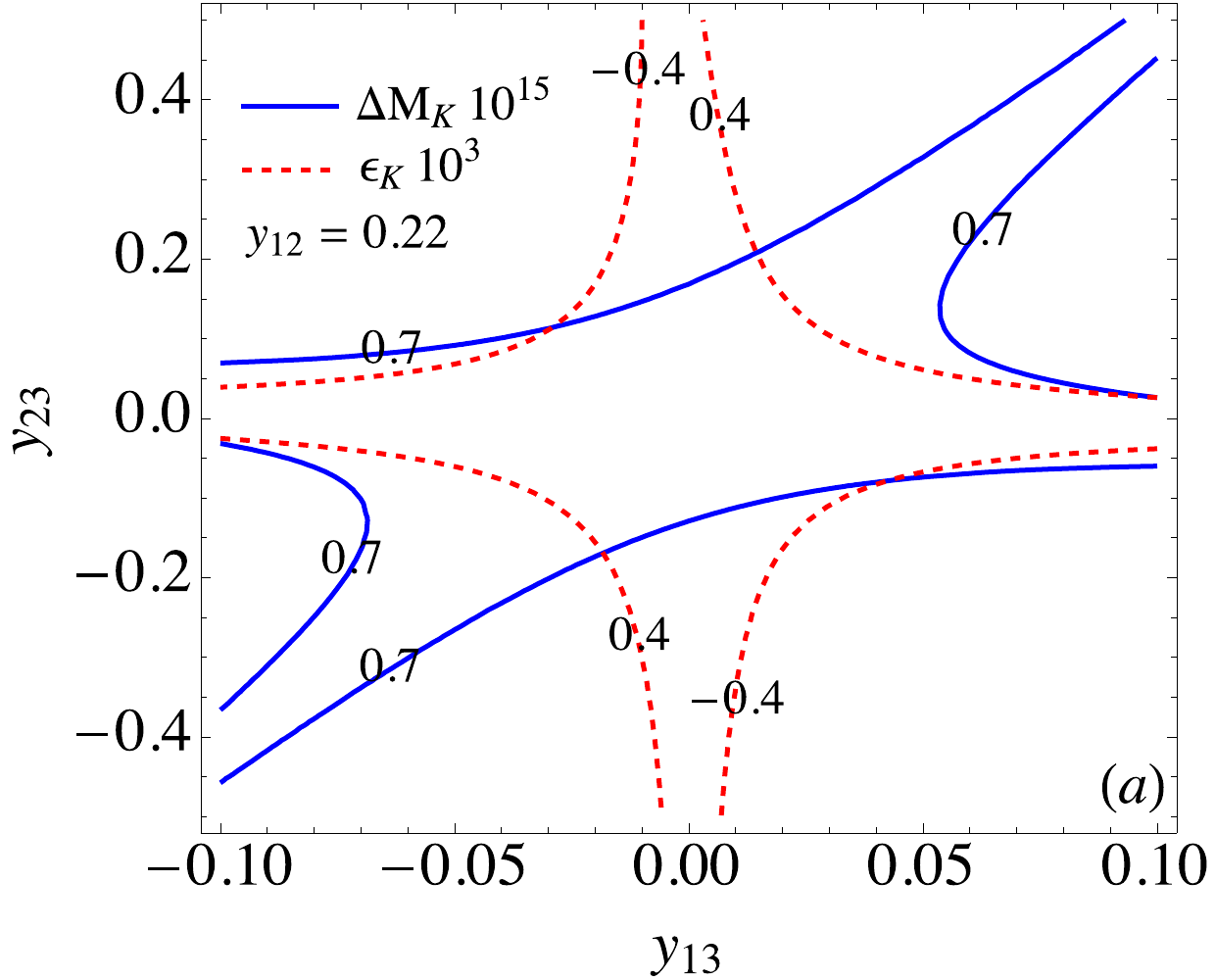}
\includegraphics[scale=0.60]{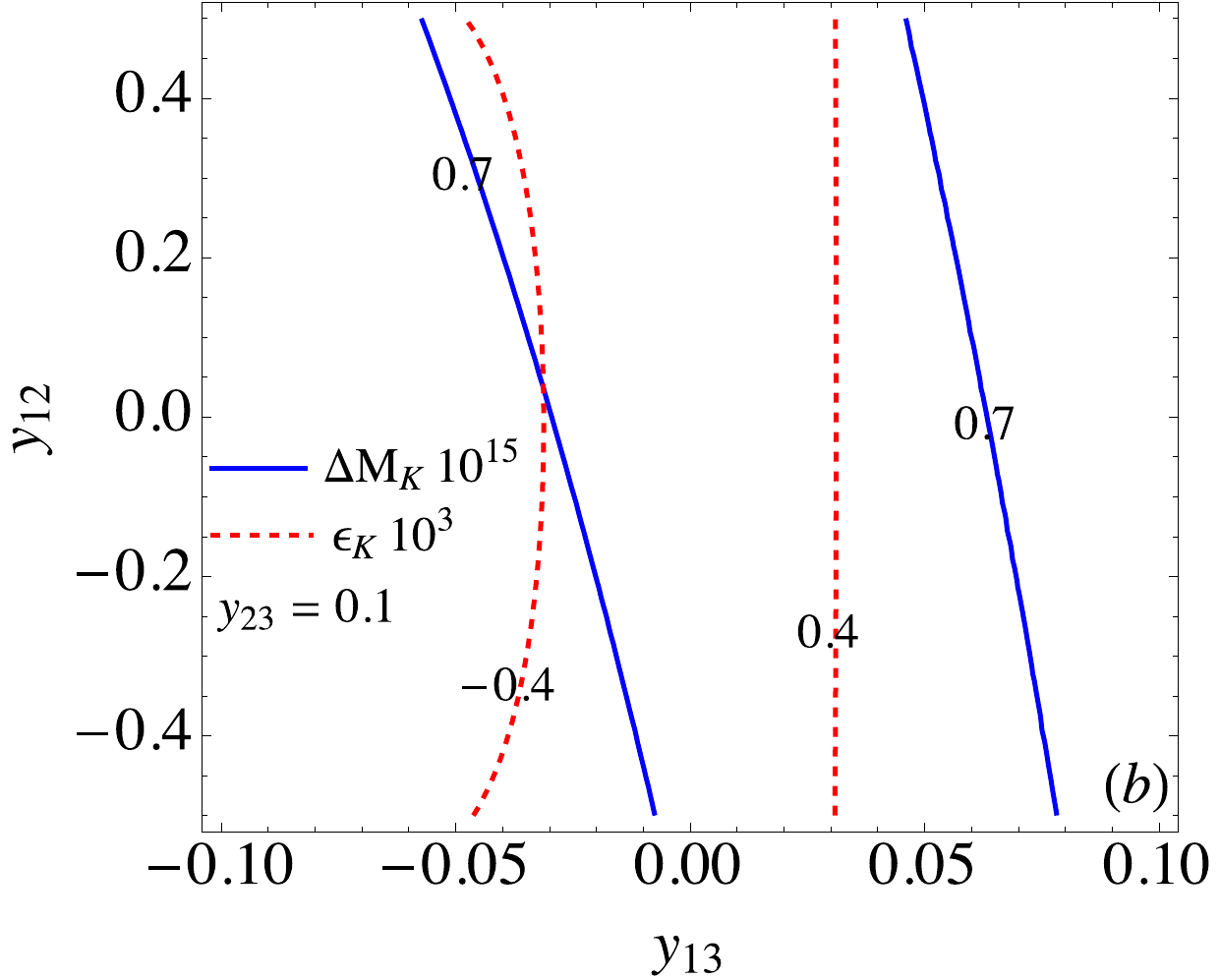}
 \caption{Contours for $\Delta M^{ H_6}_K$ (in units of $10^{-15}$) and $\epsilon^{H_{6}}_K$ (in units of $10^{-3}$), where $y_{12}=0.22$ is fixed in (a) and $y_{23}=0.1$ is fixed in (b).}
\label{fig:con_scan_ab}
\end{figure}

Although $\Delta M_K$ and $\epsilon_K$ are associated with the $({\bf y^\dagger y})_{ij}$ parameters, if we take the new Yukawa couplings to be real numbers, besides the color-sextet diquark mass, there are only three independent couplings, i.e., $y_{12}$, $y_{13}$, and $y_{23}$. In order to understand how these parameters correlate to $\Delta M_K$ and $\epsilon_K$, we show the contours of $\Delta M^{H_6}_K$ (solid) and $\epsilon^{H_6}_K$ (dashed) in Fig.~\ref{fig:con_scan_ab}(a) and (b), where we fixed $y_{12}=0.22$ and $y_{23}=0.1$ in plots (a) and (b), respectively. It can be seen that the allowed $y_{13}$ is relatively smaller than $y_{12, 23}$.

\begin{figure}[phtb]
\includegraphics[scale=0.60]{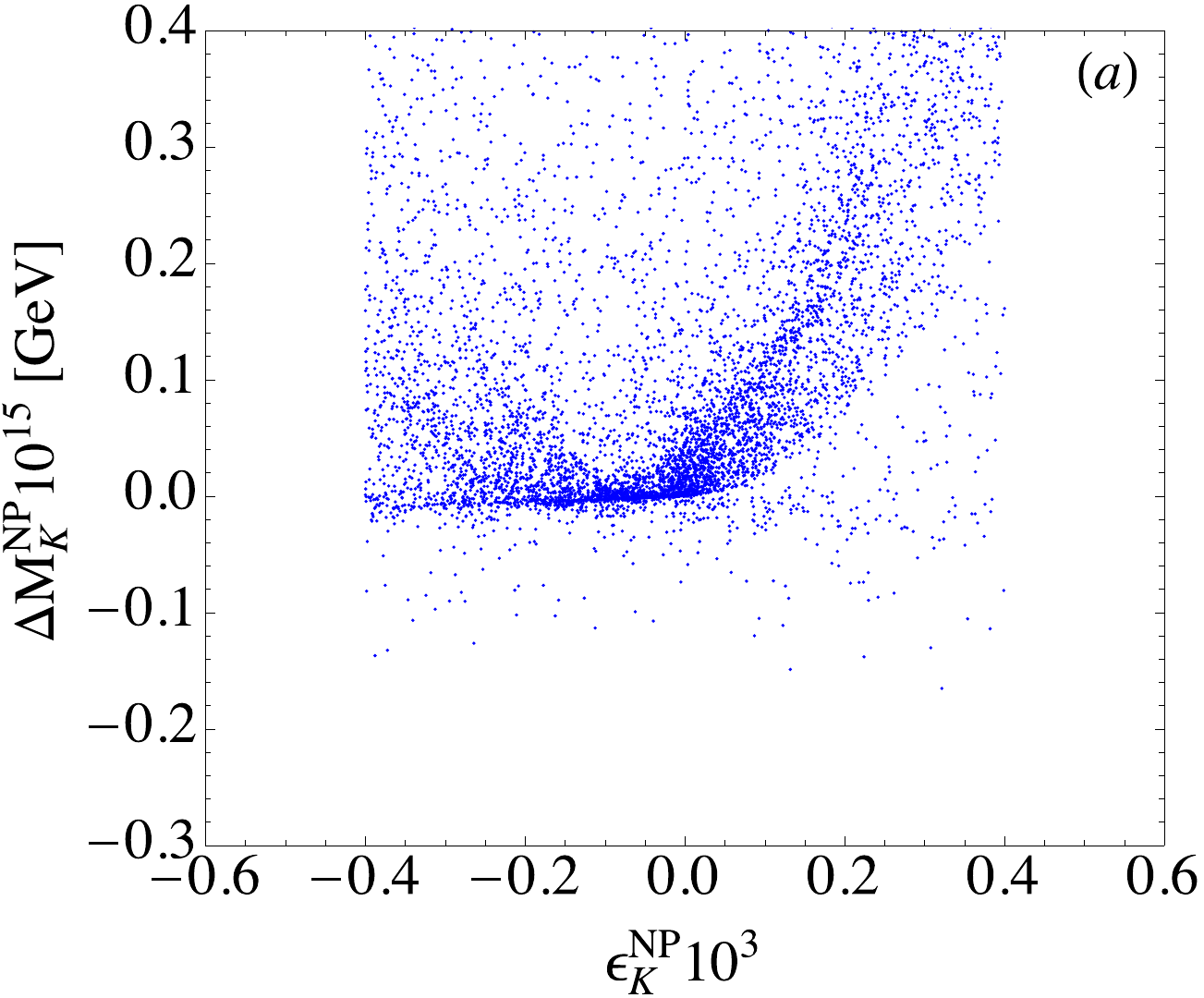}
\includegraphics[scale=0.60]{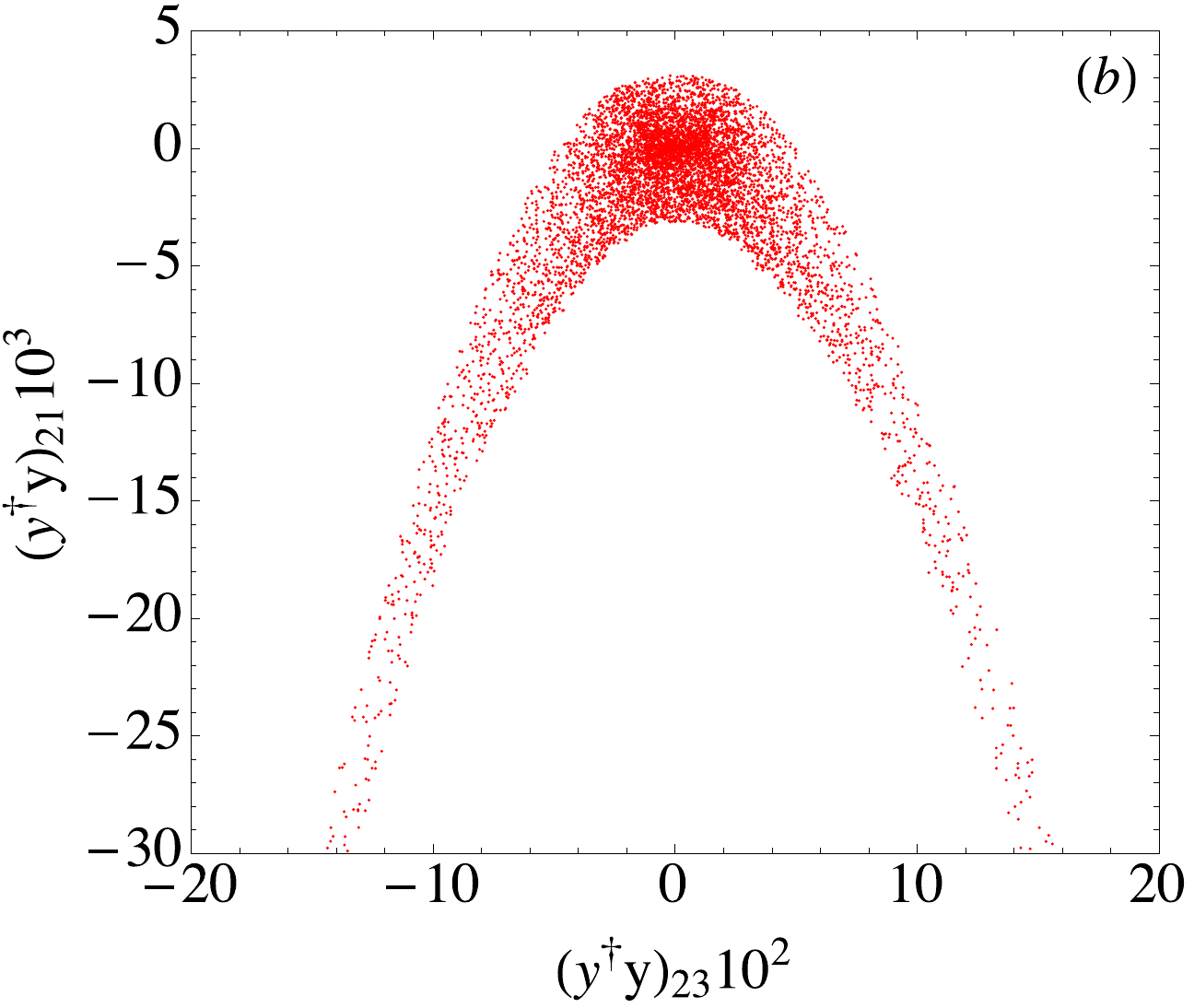}
 \caption{(a) Scatter plot for showing the allowed data points in terms of  the correlation between $\epsilon^{H_6}_{K}$ (in units of $10^{-3}$) and $\Delta M^{H_6}_K$ (in units of $10^{-15}$). (b) Scatter plot for the correlation between  $({\bf y^\dagger y})_{23}$ and $({\bf y^\dagger y})_{21}$ when the conditions of $|\Delta M^{H_6}_K| \leq  0.2 \Delta M^{\rm exp}_{K}$ and $|\epsilon^{H_6}_{K}| \leq 0.4 \times 10^{-3}$ are satisfied.}
\label{fig:limit_ab}
\end{figure}

Since $y_{12}$, $y_{13}$ and $y_{23}$ are all involved in $\Delta M_K$ and $\epsilon_K$,  to include all their effects, we scan the $y_{ij}$ parameters in the chosen regions:
 \begin{equation}
 y_{12, 13, 23} =(-0.5,\, 0.5)\,. \label{eq:para_space}
 \end{equation}
Using $3\times 10^{5}$  sampling data points,  the scatter plot for the correlation between $\Delta M^{H_6}_K$ (in units of $10^{-15})$ and $\epsilon^{H_6}_K$ (in units of $10^{-3})$ is shown in Fig.~\ref{fig:limit_ab}(a), where the  ${\cal H}^{H_6 H_6}_{\rm Box}$ and ${\cal H}^{W+G}_{\rm Box}$ effects are combined and the conditions in Eq.~(\ref{eq:conditions}) are used.  It can be seen that  there are plenty of data points remained when  the constraints from the rare $\Delta S=2$ processes are included. We will show that $\epsilon'/\epsilon$ and $K^+ \to \pi^+ \nu \bar\nu$ in the model only depend on $({\bf  y^\dagger y})_{23}$ and $({\bf  y^\dagger y})_{21}$, respectively.  From $({\bf y^\dagger y})_{21}=y^*_{32}y_{31}=y^*_{23} y_{13}$ and $({\bf  y^\dagger y})_{23} = y^*_{12} y_{13}$, the correlation between $({\bf  y^\dagger y})_{23}$ (in units of $10^{-2}$) and $({\bf  y^\dagger y})_{21}$ (in unit of $10^{-3}$) is shown in~Fig.~\ref{fig:limit_ab}(b), where the allowed data points shown in Fig.~\ref{fig:limit_ab}(a) are applied. From the plot, we clearly see that  $|({\bf  y^\dagger y})_{23}| > 0.05$ is available, and its maximum value could reach 0.16.

\section{ $\epsilon'/\epsilon$, $K\to \pi \nu  \bar\nu$, and numerical analysis}

\begin{figure}[phtb]
\includegraphics[scale=0.70]{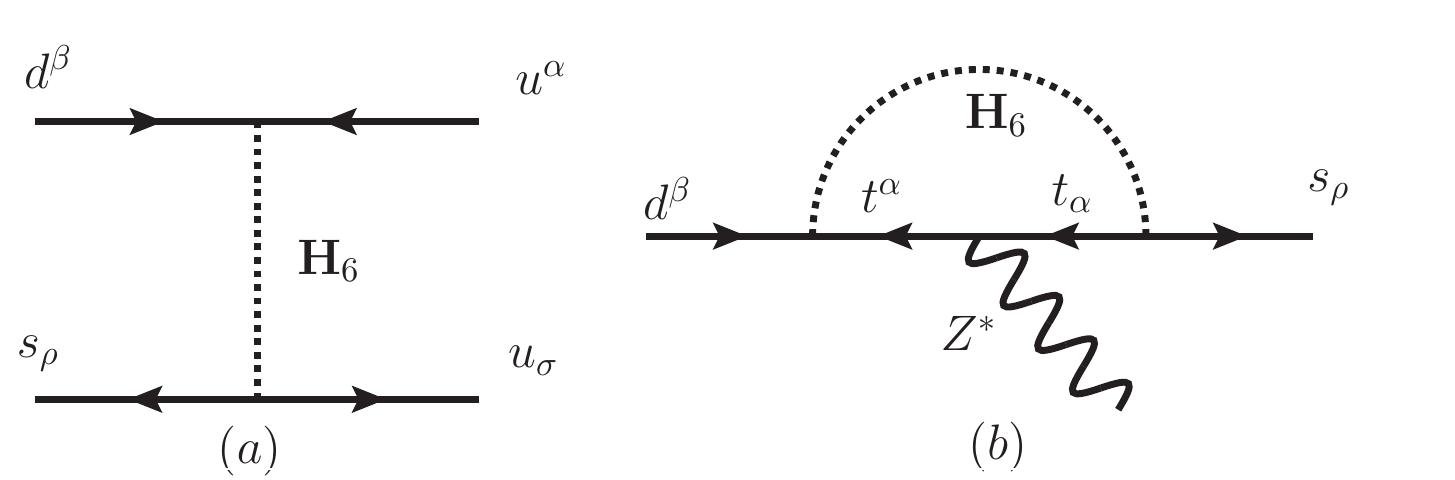}
 \caption{  (a) ${\bf H}_{6}$-mediated Feynman diagram for the $d\to s u \bar u$ decay. (b) $Z$-penguin diagram for $d\to s Z^*$ through the intermediate of ${\bf H}_6$. }
\label{fig:DS=1}
\end{figure}

In this section, we study the ${\bf H}_6$ contributions to the Kaon direct CP violation and to the rare $K\to \pi \nu \bar\nu$ decays, where the relevant Feynman diagrams are sketched in Fig.~\ref{fig:DS=1}, and the SM predictions are taken as:
 \begin{align}
 Re(\epsilon'/\epsilon)&= \left\{
\begin{array}{c}
 (1.38\pm 6.90)\times 10^{-4} ~~(\text{RBC-UKQCD~\cite{Blum:2015ywa,Bai:2015nea}})    \\
(1.9 \pm 4.5)\times 10^{-4} ~~~~~~( \text{DQCD~\cite{Buras:2015xba,Buras:2015yba,Buras:2016fys}})      
\end{array}\right. \nonumber \\
BR(K^+\to \pi^+ \nu \bar\nu) & = (8.5^{+1.0}_{-1.2})\times 10^{-11} ~~~~~\text{\cite{Bobeth:2016llm}}\,.
 \end{align}
The current experimental data are $Re(\epsilon'/\epsilon)^{\rm exp} =(16.6 \pm 2.3)\times 10^{-4}$ and $BR(K^+ \to \pi^+ \nu \bar\nu)^{\rm exp}=(1.7 \pm 1.1)\times 10^{-10}$~\cite{PDG}. Since the SM prediction on $\epsilon'/\epsilon$ is inconsistent with the experimental data by a $2\sigma$ deviation, studies of the anomaly using the new physics effects can be found in~\cite{Buras:2015qea,Buras:2015yca,Buras:2015kwd,Buras:2015jaq,Tanimoto:2016yfy,Buras:2016dxz,Kitahara:2016otd,Endo:2016aws,Bobeth:2016llm,Cirigliano:2016yhc,Endo:2016tnu,Bobeth:2017xry,Crivellin:2017gks,Bobeth:2017ecx,Haba:2018byj,Buras:2018lgu,Chen:2018ytc,Chen:2018vog,Matsuzaki:2018jui,Haba:2018rzf,Aebischer:2018rrz,Aebischer:2018quc,Aebischer:2018csl,Chen:2018dfc}.

 In addition to the tree Feynman diagram, the $K\to \pi\pi$ decays can be induced from the $Z$ penguins. Since the involved Yukawa couplings for the penguin diagrams are associated with $g^*_{i 2} g_{i1}$, in which $g^*_{12} g_{11}$ is the same as that appearing at the tree level, and $g^*_{32} g_{31}$ is dominated by the real part, it is expected that the $Z$-penguin contribution to $\epsilon'/\epsilon$ will be  smaller than  the tree Feynman diagram; therefore, we neglect the loop contributions to $\epsilon'/\epsilon$. 

Because neutrinos only couple to the $Z$ boson,  the $Z$-penguin diagram can have a significant influence on the rare $K \to \pi \nu \bar\nu$ decays. Since light quarks with $m_q \approx 0$ have no contributions to the $Z$-penguin, we only need to consider the top-quark contribution to $d\to s Z^*$. As a result, the involved Yukawa coupling is $g^*_{32} g_{31}\approx 4 ({\bf y^\dagger y})_{21}$ and  can only have a sizable influence on the CP-conserving $K^+ \to \pi^+ \nu \bar\nu$ decay. Hence, in the following analysis, we numerically show the ${\bf H}_6$ contributions to $\epsilon'/\epsilon$ and $K^+ \to \pi^+ \nu \bar\nu$, which arise from Fig.~\ref{fig:DS=1}(a) and (b), respectively.

 \subsection{ $\epsilon'/\epsilon$ from the tree diagram}
 
 Based on the Yukawa interactions in Eq.~(\ref{eq:LY}), the effective Hamiltonian for $d\to s u\bar u$ can be expressed as:
 \begin{equation}
 {\cal H}_{H_6}(d\to s u \bar u) = - \frac{ G_F V^*_{ts} V_{td}}{\sqrt{2}} C^T_{H_6} ( Q_1 + Q_2)\,, \label{eq:H_Kpipi}
 \end{equation}
where the effective Wilson coefficient $C^T_{H_6}$ at the $\mu=m_{H_6}$ scale and  the effective operators are given as:
 \begin{align}
 C^T_{H_6} & = \frac{y_W}{2} \frac{g^*_{12} g_{11} }{g^2 V^*_{ts} V_{td}}\,, \\
 Q_1 & =( \bar s d)_{V-A} (\bar u u)_{V-A}\,, \nonumber \\
  Q_2 &= ( \bar s u)_{V-A} (\bar u d)_{V-A}\,. \nonumber 
 \end{align}
Using ${\bf g}= 2 {\bf y} V$, we obtain:
 \begin{equation}
 g^*_{12} g_{11} = |y_{12} |^2 V^*_{cs} V_{cd} +|y_{13}|^2 V^*_{ts} V_{td}+ ({\bf y^\dagger y})_{23} V^*_{cs} V_{td}+({\bf y^\dagger y})_{23} V^*_{ts} V_{cd}\,.
 \end{equation}
It can be seen that if we  drop the small $Im(V^*_{ts} V_{td})$ effects, the sizable CP violating effect arises from $({\bf y^\dagger y})_{23} V^*_{cs} V_{td}$. 

The direct CP violating parameter from new physics in $K$ system can be estimated by~\cite{Buras:2015yba}:
 \begin{equation}
 Re\left( \frac{\epsilon'}{\epsilon}\right) \approx - \frac{  \omega }{\sqrt{2} |\epsilon_K|} \left[ \frac{Im A_0}{ Re A_0} - \frac{Im A_2}{ Re A_2}\right]\,,  \label{eq:epsilon_p}
  \end{equation}
 where $\omega = Re A_2/Re A_0 \approx 1/22.35$ denotes the $\Delta I=1/2$ rule, and Eq.~(\ref{eq:epsilon_p}) is only related to the ratios of hadronic matrix elements. In order to consider the hadronic effects, we employ the SM results  as~\cite{Buras:2015yba}:
 \begin{align}
ReA^{\rm SM}_0 &\approx  \frac{G_F V^*_{us} V_{ud}}{\sqrt{2}} z_{-} \langle Q_{-} \rangle_0 \left(1 + q_T \right)  \,, \nonumber \\
ReA^{\rm SM}_2 &\approx \frac{G_F V^*_{us} V_{ud}}{\sqrt{2}} z_{+} \langle Q_{+} \rangle_2 \,,
\end{align}
where the operators  $Q_+$ and $Q_-$ are defined as:
\begin{equation}
Q_{+}=\frac{1}{2} \left( Q_2 + Q_1 \right)\,, ~ Q_{-} = \frac{1}{2} \left( Q_2 - Q_1 \right)\,;
\end{equation}
$q_T= z_{+} \langle Q_{+} \rangle_0/( z_{-} \langle Q_{-} \rangle_0 )\lesssim 0.1$~\cite{Buras:2015yba};  $z_{\pm } =z_{2} \pm z_{1} $,  and the values of $z_{1,2}$ at $\mu=m_c$ are $z_1=-0.409$ and $z_2=1.212$. 

According to Eq.~(\ref{eq:H_Kpipi}), the isospin decay amplitudes for $K\to \pi\pi$ through the intermediate of ${\bf H_6}$ can be written as:
 \begin{equation}
 A^{H_6}_{0(2)} = - \frac{ G_F V^*_{ts} V_{td}}{\sqrt{2}} C^T_{H_6} 2 \langle Q_+ \rangle_{0(2)}\,.
 \end{equation}
 If we take $ReA^{\rm SM}_{0(2)}\approx ReA^{\rm exp}_{0(2)}$ from Eq.~(\ref{eq:epsilon_p}), the Kaon direct CP violation can then be expressed as:
\begin{align}
 Re\left( \frac{\epsilon'}{\epsilon}\right)^T_{H_6} & \approx  T^{(1/2)}_{H_6} -  T^{(3/2)}_{H_6}\, \\
 T^{(1/2)}_{H_6} & =0.705 \frac{r_1\, y_W\, q_T }{z_{+} (1 + q_T)} Im\left(\frac{4 ({\bf y^\dagger y})_{23} V_{td}}{g^2}\right)\,, \nonumber \\
 T^{(3/2)}_{H_6} & =0.705 \frac{r_1\, y_W  }{z_{+} } Im\left(\frac{4 ({\bf y^\dagger y})_{23} V_{td}}{g^2}\right)\,,
  \end{align}
where the factor of $0.705$ is the renormalization group running effect from $\mu=m_{H_6}$ to $\mu = m_c$. Due to the small $q_T$ factor, the $T^{(1/2)}_{H_6}$ contribution is  smaller than  $T^{(3/2)}_{H_6}$. Since $Re(\epsilon'/\epsilon)^{T}_{H_6}$ is linearly proportional to $({\bf y^\dagger y})_{23}$, we show the values of $Re(\epsilon'/\epsilon)^{T}_{H_6}$ (in units of $10^{-3}$) with the benchmarks referenced in Table~\ref{tab:ep/e}. It can be seen that the results can be as large as the SM prediction. 

\begin{table}[hpbt]
\caption{ $Re(\epsilon'/\epsilon)^T_{H_6}$ with the taken benchmarks of $({\bf y^\dagger y})_{23}$. }
\label{tab:ep/e}
\begin{tabular}{c|cccccc} \hline \hline
 
 $({\bf y^\dagger y})_{23}$~~~&~~~$0.05$~~~&~~~$0.07$~~~&~~~$0.09$~~~&~~~ $0.12$~~~&~~~ $0. 14$~~~&~~~$0. 16$  \\ \hline
 $Re(\epsilon'/\epsilon)^{T}_{H_6}(10^{-3})$  & $0.25$  & $0.34$ & ~$0.44$ & ~~$0.59$ &  ~~$0.69$ & ~~$0.79$\\ \hline\hline
\end{tabular}
\end{table}

\begin{figure}[phtb]
\includegraphics[scale=0.60]{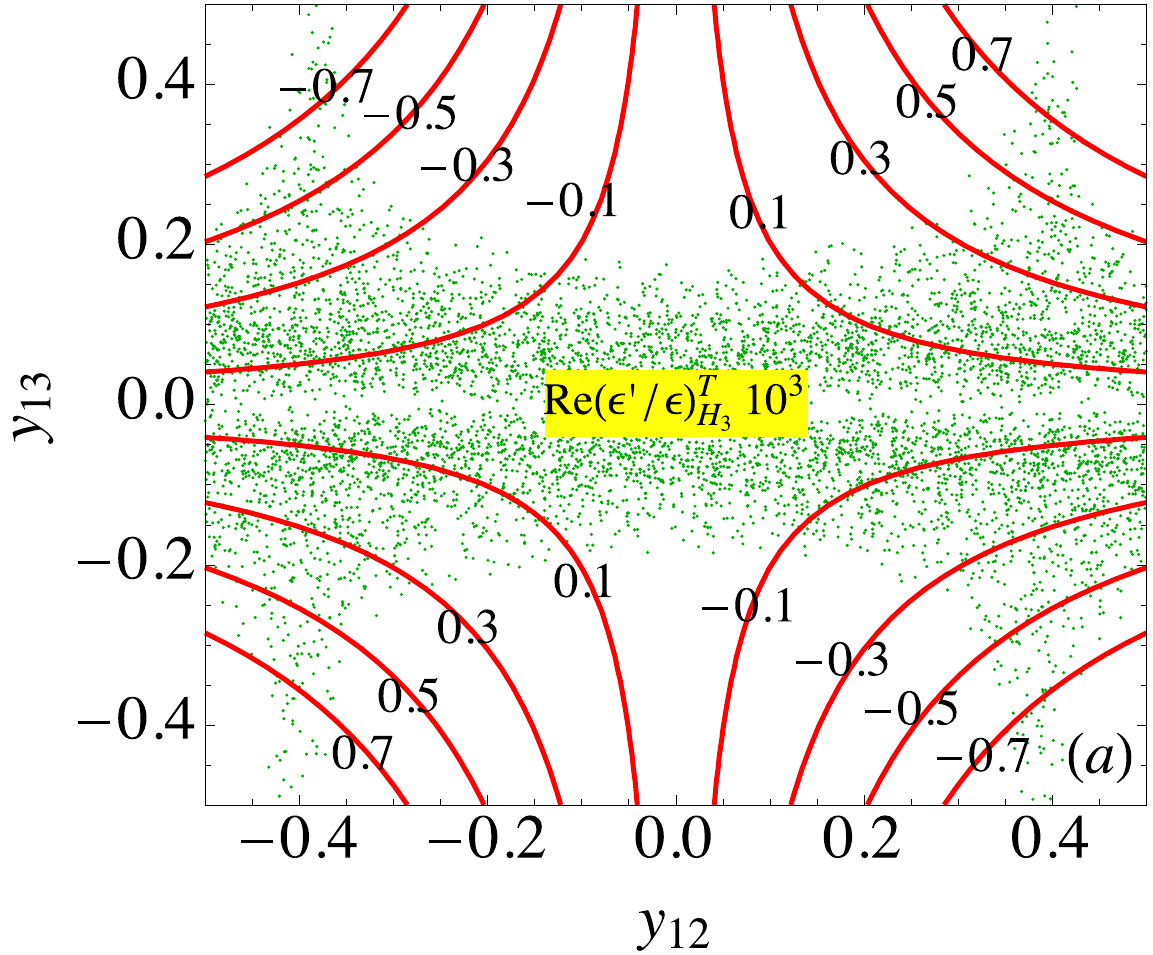}
\includegraphics[scale=0.60]{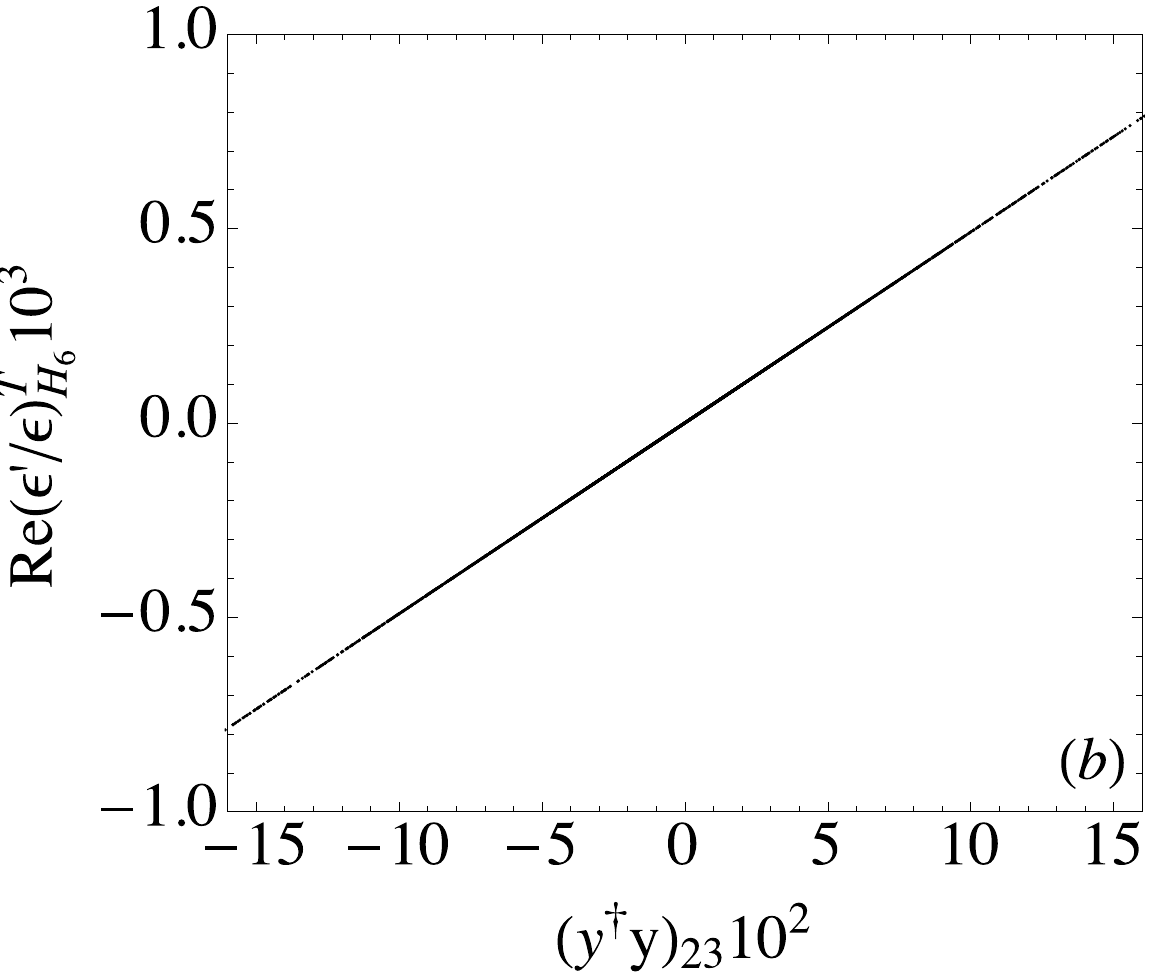}
 \caption{ (a) Contours for $Re(\epsilon'/\epsilon)^T_{H_6}$  (in units of $10^{-3}$)  as a function of $y_{12}$ and $y_{13}$, where the scatters are the allowed data points.  (b) Scatter plot for $(\epsilon'/\epsilon)^T_{H_6}$with respect to $({\bf  y^\dagger y})_{23}$ (in units of $10^{-2}$).}
\label{fig:epoveK_yy23}
\end{figure}

According to  the results shown in Fig.~\ref{fig:limit_ab}(b), it can be found that the upper value of $|({\bf y^\dagger y})_{23}|$ can reach $0.16$. From Table~\ref{tab:ep/e}, it is expected  that the maximum value of $\epsilon'/\epsilon$ in the model can be around 
 $0.79 \times 10^{-3}$. In order to understand how  $\epsilon'/\epsilon$ correlates to the relevant parameters, using $({\bf y^\dagger y})_{23}= y^*_{12} y_{13}$, the contours of $Re(\epsilon'/\epsilon)^{T}_{H_6}$ (in units of $10^{-3}$) as a function of $y_{12}$ and $y_{13}$ are shown in Fig.~\ref{fig:epoveK_yy23}(a), where the scatters denote the allowed data points of which the required conditions in Eq.~(\ref{eq:conditions}) are satisfied.  Moreover, using the same allowed data points,  the scatter plot for $Re(\epsilon'/\epsilon)^T_{H_6}$ with respect to  $({\bf y^\dagger y})_{23}$ is shown in Fig.~\ref{fig:epoveK_yy23}(b), where the results match the values in Table~\ref{tab:ep/e}. In the following analysis, we demonstrate that the allowed $({\bf  y^\dagger y})_{23}$ region will be further bounded by the $K^+\to \pi^+ \nu \bar\nu$ decay.

 \subsection{$K\to \pi \nu \bar\nu$ from the $Z$-penguin}
 
 Using the $d\to s Z^*$ result, which is induced from the color-triplet quark and was obtained in~\cite{Chen:2018dfc}, the Lagrangian for $d\to sZ^*$ through the ${\bf H}_6$ loop can be written as:
 \begin{equation}
 {\cal L}_{d\to s Z^*} = - \frac{g}{\cos\theta_W} C^{sd}_Z  \bar s \gamma_\mu P_L d \, Z^\mu+ H.c.\,, 
 \end{equation}
where $\theta_W$ is the Weinberg's angle and $C^{sd}_Z$ is given as:
 \begin{align}
C^{sd}_Z & = \frac{ g^*_{32} g_{31}}{(4\pi)^2} I_Z(y_t)\,, \nonumber \\
I_{Z}(y_t) & = - \frac{y_t}{1-y_t} - \frac{y_t \ln y_t }{(1-y_t)^2}\,. 
 \end{align}
According to the SM $Z$-boson coupling to the neutrinos, the effective Hamiltonian for $d\to s \nu \bar\nu$ can be expressed as:
 \begin{align}
 {\cal H}^{H_6}_{d\to s \nu \bar\nu} & =  \frac{G_F \lambda^{sd}_t}{\sqrt{2}} \frac{\alpha}{\pi}  C^{ds}_{\nu}  \bar s \gamma_\mu P_L d\, \bar \nu \gamma^\mu(1-\gamma_5) \nu+ H.c.\,, \nonumber \\
C^{sd}_\nu  & = \frac{X^{{ H}_6}_t }{\sin^2\theta_W}  =    \frac{1 } {2\sin^2\theta_W} \frac{g^*_{32} g_{31} }{g^2 \lambda^{sd}_t  }I_{Z} (y_t)  \,,
 \end{align}
with $\lambda^{ds}_{q'} = V^*_{q' s} V_{q' d}$.

\begin{figure}[phtb]
\includegraphics[scale=0.60]{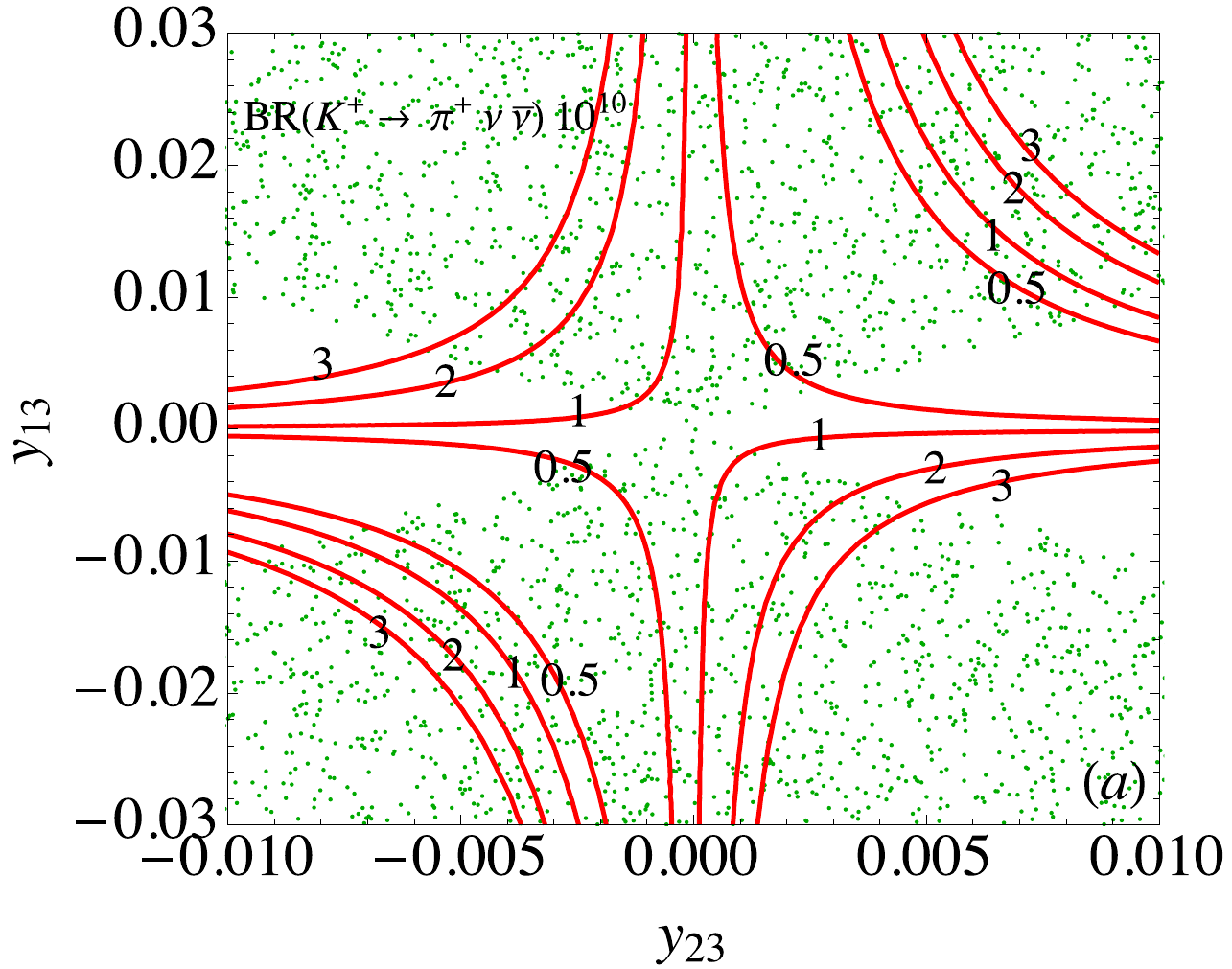}
\includegraphics[scale=0.60]{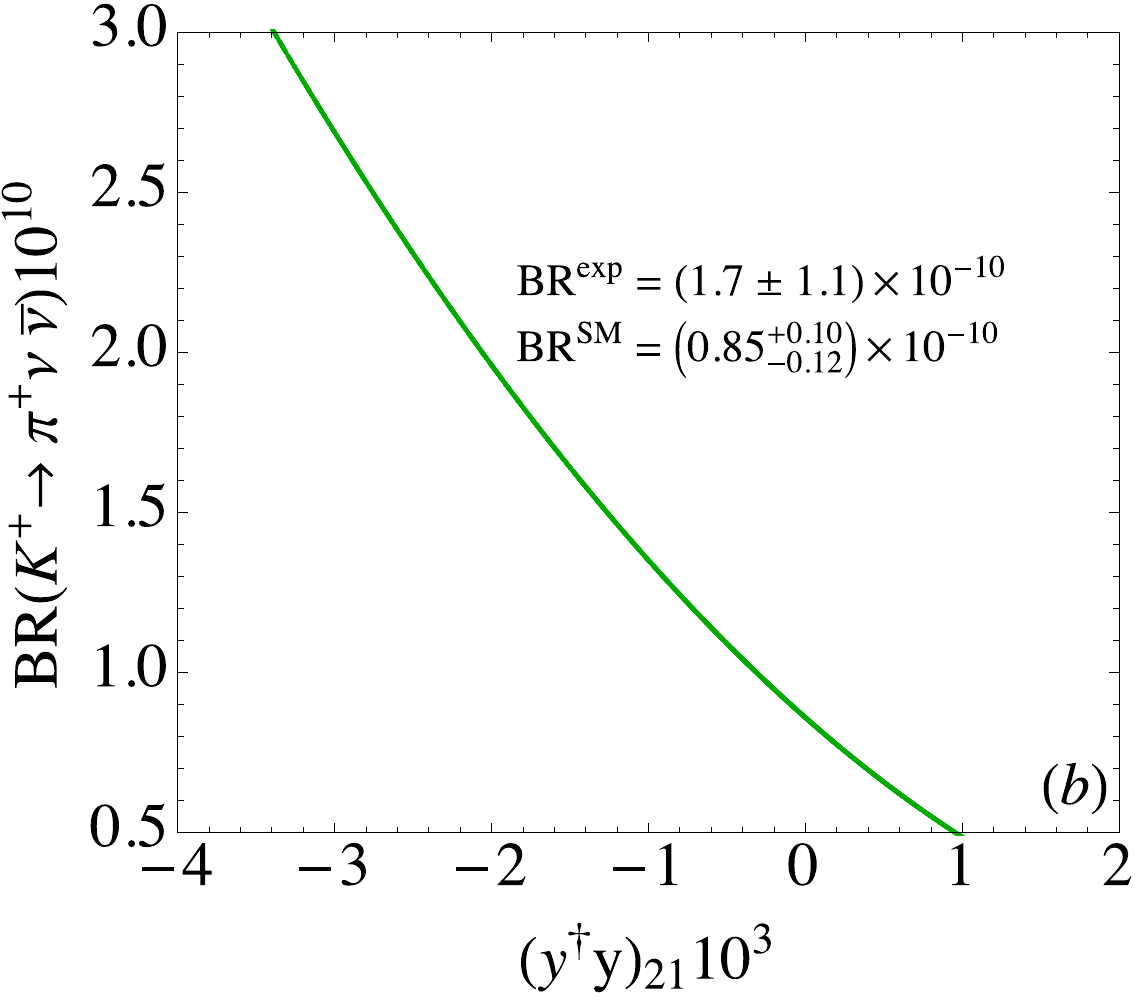}
 \caption{(a) Contours for $BR(K^+\to \pi^+ \nu \bar\nu)$  (in units of $10^{-10}$) as a function of $y_{23}$ and $y_{13}$, where the scatters denote the data points constrained by the $\Delta S=2$ processes. (b) Scatter plot for   $BR(K^+\to \pi^+ \nu \bar\nu)$ with respect to $({\bf y^\dagger y})_{21}$ (in units of $10^{-3}$). }
\label{fig:BRKpinunu}
\end{figure}

 Based on the formula shown in~\cite{Bobeth:2017ecx}, the BR for $K^+ \to \pi^+ \nu \bar\nu$ from the ${\bf H}_{6}$ and SM contributions can be formulated as:
 \begin{equation}
 BR(K^+ \to \pi^+ \nu \bar\nu) = \frac{\kappa_+ (1+\Delta_{\rm EM})}{\lambda^{10}} \left[ |Re\left( \lambda^{sd}_c X_c + \lambda^{ds}_t X_t \right)|^2 + |Im\left( \lambda^{ds}_t X_t\right)|^2\right]\,, 
 %
 \end{equation}
where $\Delta_{EM} = -0.003$; $X_c= \lambda^4 P_c(X)\approx 0.404 \, \lambda^4$ denotes the charm-quark contribution~\cite{Isidori:2005xm,Mescia:2007kn}; $X_t = X^{\rm SM}_t + X^{{\rm H}_6}_t$ with $X^{\rm SM}_t\approx 1.481$, and $\kappa\approx 5.173 \times 10^{-11}$~\cite{Buras:2015qea}. Using the allowed data points, which were earlier obtained by scanning $y_{ij}$ parameters,  we show the contours for $BR(K^+ \to \pi^+ \nu \bar \nu)$ (in units of $10^{-10}$) as a function of $y_{23}$ and $y_{13}$ in Fig.~\ref{fig:BRKpinunu}(a), where the scatters denote the allowed data points. In addition, we also show $BR(K^+ \to \pi^+ \nu \bar \nu)$ with respect to $({\bf y^\dagger y})_{21}$  in Fig.~\ref{fig:BRKpinunu}(b). Since a negative $({\bf y^\dagger y})_{21}$ leads to constructively interfere with the SM contribution due to $V_{ts} <0$, it can be seen that $BR(K^+ \to \pi^+ \nu \bar\nu)$ is sensitive to the negative $({\bf y^\dagger y})_{21}$ and it can reach the value of current experimental measurement. From the plots,  it can be seen  that the resulting $BR(K^+ \to \pi^+ \nu \bar\nu)$ can not only reach the current upper limit but also  can further exclude the parameter space, e.g., the region of $({\bf y^\dagger y})_{21}< -3.1 \times 10^{-3}$. 


\begin{figure}[phtb]
\includegraphics[scale=0.60]{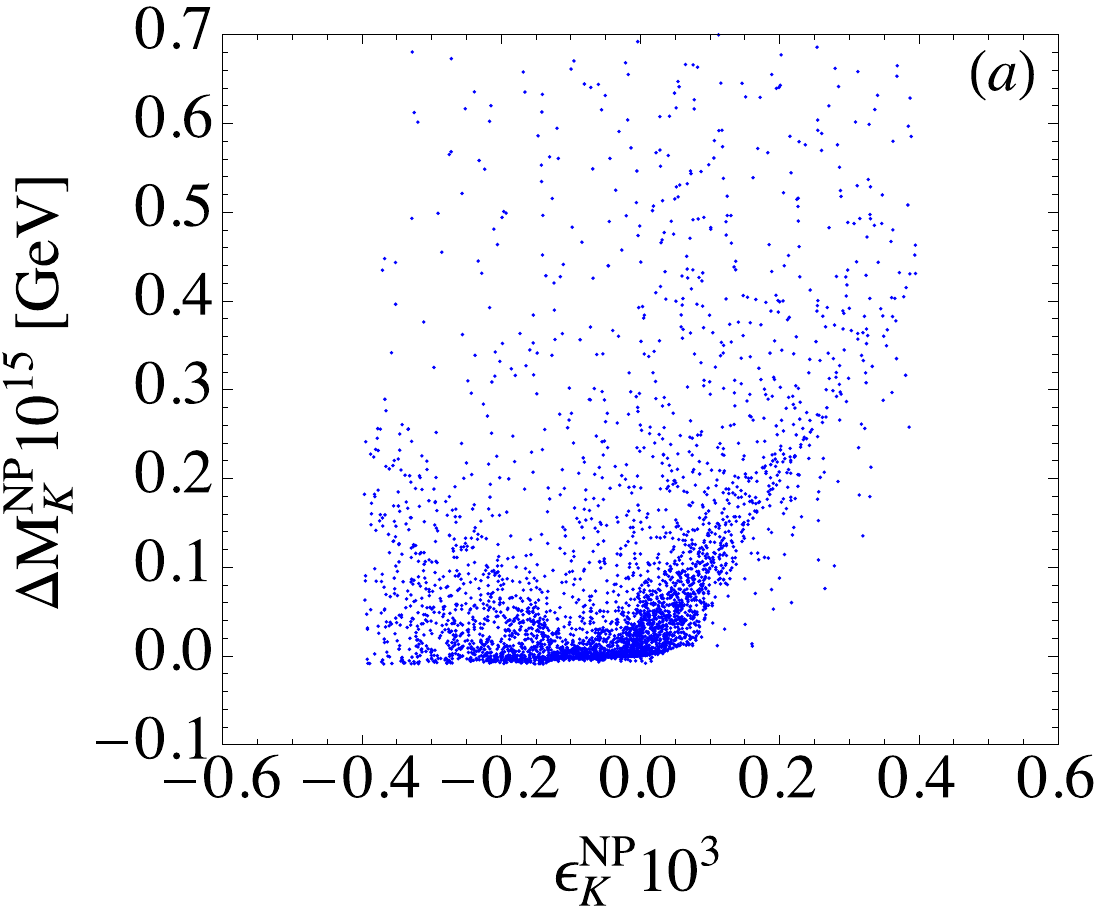}
\includegraphics[scale=0.60]{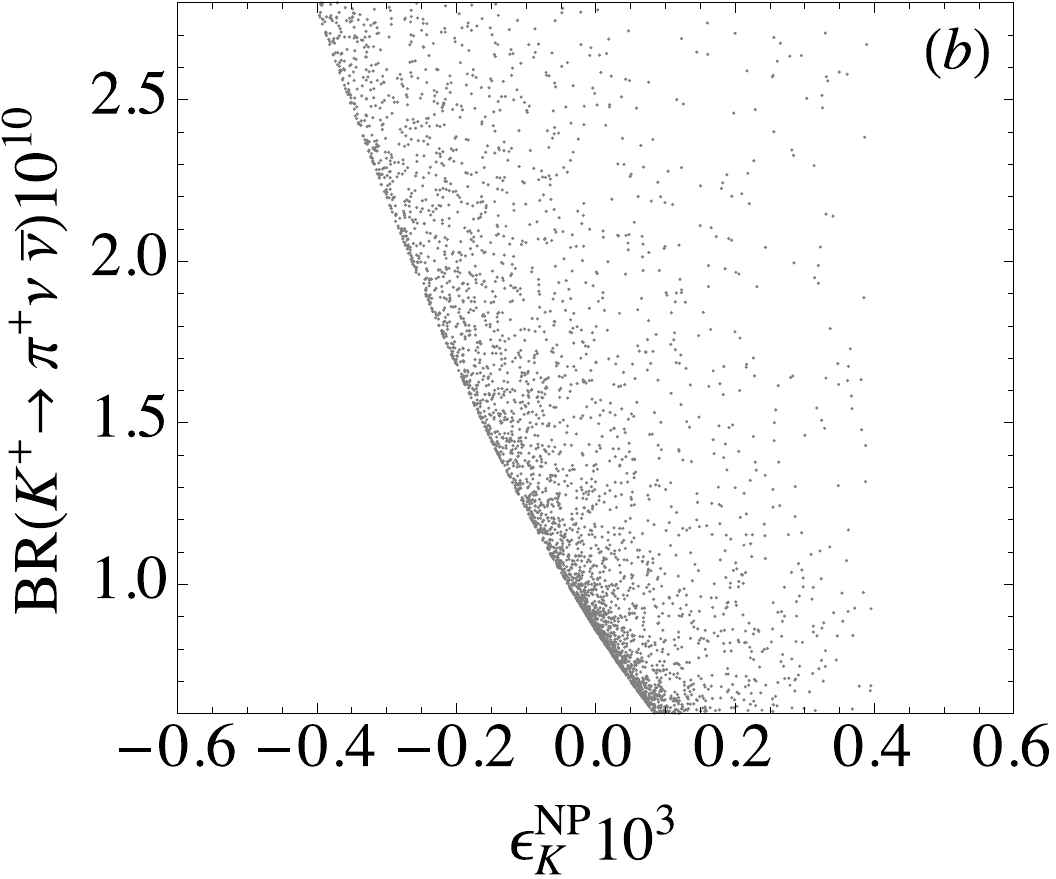}
\includegraphics[scale=0.60]{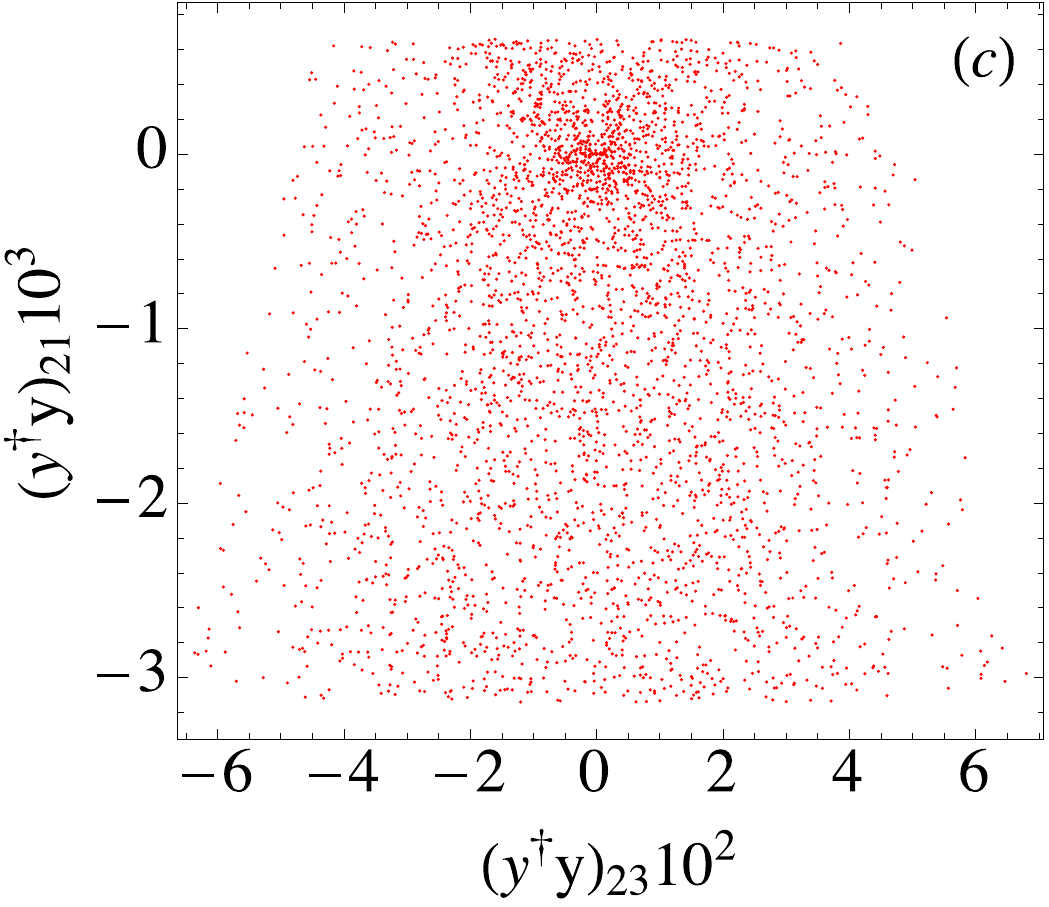}
\includegraphics[scale=0.60]{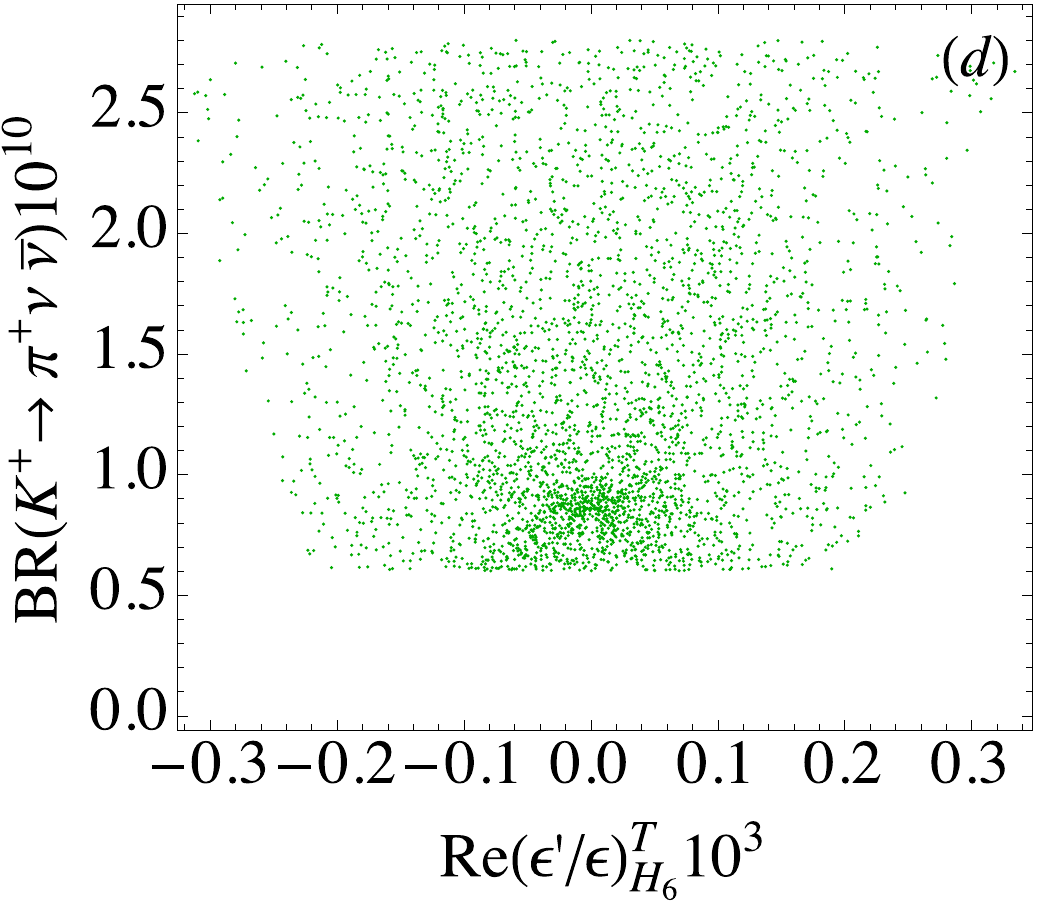}
 \caption{Scatter plot for (a) $\Delta M^{\rm NP}_K$ vs  $\epsilon^{\rm NP}_K$; (b) $BR(K^+ \to \pi^+ \nu \bar\nu)$ vs $\epsilon^{\rm NP}_K$; (c) $({\bf y^\dagger y})_{21}$ vs $({\bf y^\dagger y})_{23}$, and (d)  $BR(K^+ \to  \pi^+ \nu \bar\nu)$ vs $Re(\epsilon'/\epsilon)^T_{H_6}$. }
\label{fig:scan_final}
\end{figure}

To clearly show the influence when the  constraints from  $\epsilon_K$, $\Delta M_K$, and $BR(K^+ \to \pi^+ \nu \bar\nu)$ are combined together, we rescan the $y_{12,13,23}$ parameters using $3\times 10^{5}$ sampling points. 
 The results are shown in Fig.~\ref{fig:scan_final}. Comparing to Fig.~\ref{fig:limit_ab}, it can be seen that the allowed data points are reduced and  the allowed regions of $({\bf y^\dagger y})_{21,23}$ are shrunk  due to $BR(K^+ \to \pi^+ \nu \bar\nu)$.  As a result, the maximum value of the direct Kaon CP violation in the model becomes $Re(\epsilon'/\epsilon)^T_{H_6} \sim 0.3 \times 10^{-3}$ while the BR for $K^+ \to \pi^+ \nu \bar\nu$ can still fit the current data. 

\section{Summary}

We studied the left-handed diquark contributions to the rare $K$ processes, including $\Delta M_K$, $\epsilon_L$, $\epsilon'/\epsilon$, and $K^+ \to \pi^+ \nu \bar\nu$. We showed that the box diagrams mediated by the $W(G)$ and ${\bf H}_6$ bosons vanish when the limit of $m_t=0$ is taken and the flavor mixings $V^{u,d}_L$ are introduced. However, when the heavy top-quark mass effect is taken into account, it was found that  the nonvanished $W(G)-{\bf H}_6$ box diagrams can be significantly enhanced and can be as large as the pure ${\bf H}_6$ box diagrams.

Before including $W(G)-{\bf H}_6$ effects, due to the CKM hierarchical structure, we found that  $\Delta M_K$ gives the most strict constraints on the parameters with the exception of $({\bf y}^\dagger {\bf y})_{23}$, which the $\epsilon_K$ gives the strongest constraint. However, when the sizable $W(G)-{\bf H}_6$ effects are taken into account,  the constraints on $({\bf y}^\dagger {\bf y})_{21}$ from $\Delta M_K$ and  $\epsilon_K$ become comparable.  For comparison, we also analyzed the right-handed diquark effects. It turns out that the relevant couplings obtain a stricter constraint from the $\Delta S=2$ processes and are relatively smaller than the left-handed diquark couplings. 

With the bounded parameters, which include the $K^+\to \pi^+ \nu \bar\nu$ constraint,  the Kaon direct CP violation $\epsilon'/\epsilon$ mediated by the ${\bf H}_6$ at the tree level can reach $Re(\epsilon'/\epsilon)\sim 0.3 \times 10^{-3}$. Although  the ${\bf H}_6$-mediated $Z$-penguin contribution to $K_L\to \pi^0 \nu \bar\nu$ is small in the current chosen scheme, the contribution to $K^+ \to \pi^+ \nu \bar \nu$ can fit the current experimental upper bound. Hence, the NA62 experiment can further constrain the parameter space if no signals of $K^\to \pi^+ \nu \bar \nu$ are observed. However, if an unexpected large BR for $K^+ \to \pi^+ \nu \bar\nu$ is found in NA62, the diquark can be a potential candidate explaining the excess.

\appendix

\section{CP phases of the diquark Yukawa couplings}

We studied the ${\bf H}_6$ effects with the assumption of a real ${\bf y}$ matrix. However, it is of interest to examine how many physical CP phases involve in ${\bf y}$ when ${\bf y}$ is taken as a complex matrix. To explore the question, before EWSB, we can choose the quark states in a way that  the SM Higgs Yukawa couplings to the up-type quarks are diagonalized and expressed as:
 \begin{equation}
 Y^u_{\rm dia} = V^u_L Y^u V^{u\dagger}_{R}\,, \label{eq:Yu}
 \end{equation}
where $Y^u$ denotes the Yukawa matrix; $Y^u_{\rm dia}$ is a real and diagonal matrix, and $V^{u}_{L,R}$ are the unitary matrices. Since $Y^u$ is an arbitrary $3\times 3$ complex matrix, we have freedom to choose the quark basis such that $Y^u$ is diagonalized before EWSB. 

In this basis, Eq.~(\ref{eq:L_Y}) can be written as:
 \begin{align}
 -{\cal L}_Y & = u^T_L C {\bf f}  {\bf H}^\dagger_{6} d_L+ H.c.\,, \label{eq:L_Y_f}
 \end{align}
with ${\bf f}=\left( 2 V^{u*}_L {\bf y}V^{u\dagger}_L \right)$. Since the down-type quarks are not the physical eigenstates, we can rotate the down-type quark states through  the $d_L \to V^u_L d_L$ transformation; then, ${\bf f}$ is an antisymmetric matrix and can represent ${\bf y}$.   From above analysis, it can be seen that choosing a proper basis,  ${\bf y}$ is still an antisymmetric matrix when $Y^u$ is diagonalized before EWSB. In the following discussions, we use ${\bf f}$ and $Y^u_{\rm dia}$ instead of ${\bf y}$ and $Y^u$. 

Next, we analyze how many physical CP phases exist in ${\bf f}$. Set the complex elements of ${\bf f}$ as: $f_{12}=|f_{12}| e^{i\theta_{12}}$, $f_{13}=|f_{13}|e^{i\theta_{13}}$, and $f_{23}=|f_{23}|e^{i\theta_{23}}$, where $\theta_{12,13,23}$ are independent CP phases. In order to rotate the unphysical CP phases away,  we transform the quark states as $u_{L,R} \to V u_{L,R}$, and $d_L \to V d_L$ with $V(\theta)={\rm diag}(1, e^{i\theta_2}, e^{i\theta_3})$, where we remove one overall phase from $V_{11}$. Then, the $Y^u_{\rm dia}$ and ${\bf y}$ are  respectively transformed as $Y^u_{\rm dia} \to Y^u_{\rm dia}$ and 
 \begin{equation}
{\bf f}'=V^*(\theta) {\bf f} V^ \dagger(\theta)= 
\left(
\begin{array}{ccc}
 0 & f_{12}e^{-i\theta_2} &  f_{13}e^{-i\theta_3}  \\
f_{21} e^{-i\theta_2} &  0 & f_{23} e^{-i(\theta_2 + \theta_3)}     \\
 f_{31} e^{-i\theta_3 }  &  f_{32} e^{-i(\theta_2 + \theta_3)}   &   0
\end{array}
\right)\,. \label{eq:app_f}
 \end{equation}
Since $\theta_{2,3}$ are free parameters, we can rotate away the phases of $f_{12}$ and $f_{23}$ requiring  $\theta_{12} = \theta_{2}$ and  $\theta_{13} =  \theta_3$. Since $\theta_2$ and $\theta_3$ are fixed, the phase of the ${\bf f'}_{23}$ element is  $f_{23} e^{-i(\theta_2 + \theta_3)}=|f_{23}|e^{i(\theta_{23}-\theta_2 -\theta_3)}$ and cannot be rotated away. Hence, we conclude that before EWSB,  there is only one unrotated CP phase in ${\bf f}$.  

Based on the basis that leads to $Y^u_{\rm dia}$, we now consider the flavor mixings after EWSB.  We can introduce the $U^d_{L,R}$ unitary matrices to diagonalize the down-type quark Yukawa matrix as:
 \begin{equation}
 Y^d \to U^d_L Y^d U^{d\dagger}_L = Y^d_{\rm dia}\,. 
 \end{equation}
Since up-type quark Yukawa matrix has been diagonalized, we have  $U^u_{L,R}=1$. Accordingly, the CKM matrix can be expressed as $V_{\rm CKM}= U^u_L U^{d\dagger}_L = U^{d\dagger}_L$.  Thus, in terms of mass eigenstates, Eq.~(\ref{eq:L_Y_f}) can be written as:
 \begin{equation}
 -{\cal L}_Y  = u^T_L C {\bf f}  V_{\rm CKM} {\bf H}^\dagger_{6} d_L+ H.c.\,. \label{eq:L_Y_fV}
 \end{equation}
 Clearly, ${\bf f}  V_{\rm CKM}$ is not an antisymmetric matrix.
So far, $V_{\rm CKM}$ is  a generic unitary matrix and the  unphysical phases have not yet been removed. From the SM charged weak interactions, which is written as $\bar u_L \gamma_\mu V_{\rm CKM} d_L W^{+\mu}$, we can see that  five of six phases in $V_{\rm CKM}$ can be removed by redefining the phases of up- and down-type quarks. If we choose that up- and down-type quarks absorb two and three CP phases from $V_{\rm CKM}$, respectively, from Eq.~(\ref{eq:L_Y_fV}), it can be seen that the two absorbed CP phases from the up-type quarks will flow to the diquark Yukawa coupling as ${\bf f''}=V^*(\phi){\bf f} V^\dagger(\phi)$ with $V(\phi)=(1, e^{i\phi_2}, e^{i\phi_3})$, where the matrix form of ${\bf f''}$ is the same as that in Eq.~(\ref{eq:app_f}) and  has three CP phases, i.e., $\phi_2$, $\phi_3$, and $\theta_{23}-\theta_2-\theta_3$. Again, one of the three CP phases is a global phase and can be absorbed to the up-quark states.  Hence, if $V_{\rm CKM}$ in Eq.~(\ref{eq:L_Y_fV}) carries one physical CP phase, the associated ${\bf f}$ matrix in principle can have two independent CP phases. 

\section*{Acknowledgments}

This work was partially supported by the Ministry of Science and Technology of Taiwan,  
under grants MOST-106-2112-M-006-010-MY2 (CHC).

\end{document}